\pdfoutput=1
\documentclass[11pt,authoryear,a4paper]{article}
\usepackage[margin=2.5cm]{geometry}
\usepackage[T1]{fontenc}
\usepackage[english]{babel}
\usepackage[utf8x]{inputenc}
\usepackage{ucs}
\usepackage{amsmath}
\usepackage{amsfonts}
\usepackage{amssymb}
\usepackage{array}
\usepackage{graphicx}
\usepackage{subfig}
\usepackage{color, natbib}
\usepackage{authblk}
\usepackage[]{algorithm}
\usepackage{algpseudocode}
\usepackage{url}
\RequirePackage[colorlinks,linkcolor=black,citecolor=blue,urlcolor=blue]{hyperref}

\newcommand{\bC}{\mathbf{C}}
\newcommand{\bY}{\mathbf{Y}}
\newcommand{\bF}{\mathbf{F}}
\newcommand{\bG}{\mathbf{G}}
\newcommand{\bH}{\mathbf{H}}
\newcommand{\bX}{\mathbf{X}}

\newcommand{\bV}{\mathbf{V}}

\newcommand{\bS}{\mathbf{S}}
\newcommand{\bs}{\mathbf{s}}
\newcommand{\by}{\mathbf{y}}
\newcommand{\bx}{\mathbf{x}}
\newcommand{\bz}{\mathbf{z}}
\newcommand{\btheta}{\boldsymbol{\theta}}

\author{Umberto Picchini${}^{a,*}$, Adeline Samson${}^{b,*}$ \bigskip
  \\
  ${}^{a}${\small Centre for Mathematical Sciences},
  {\small S\"{o}lvegatan 18},
  {\small SE-22100 Lund, Sweden}\\
  {\small Email:} {\small {\tt umberto@maths.lth.se}}  \\  
  ${}^{b}${\small LJK},
  {\small Universite Grenoble Alpes, F-38000 Grenoble, France;}\\ 
  {\small CNRS, LJK, F-38000 Grenoble, France}\\
  {\small Email:} {\small {\tt adeline.leclercq-samson@imag.fr}}  \\
  }

\title{Coupling stochastic EM and Approximate Bayesian Computation for parameter inference in state-space models}

\date{}

\begin{document}

\maketitle
\begin{abstract}
We study the class of state-space models and perform maximum likelihood estimation for the model parameters. We consider a stochastic approximation expectation-maximization (SAEM) algorithm to maximize the likelihood function with the novelty of using approximate Bayesian computation (ABC) within SAEM. The task is to provide each iteration of SAEM with a filtered state of the system, and this is achieved using an ABC sampler for the hidden state, based on sequential Monte Carlo (SMC) methodology. It is shown that the resulting SAEM-ABC algorithm can be calibrated to return accurate inference, and in some situations it can outperform a version of SAEM incorporating the bootstrap filter. Two simulation studies are presented, first a nonlinear Gaussian state-space model then a state-space model having dynamics expressed by a stochastic differential equation.  Comparisons with iterated filtering for maximum likelihood inference, and Gibbs sampling and particle marginal methods for Bayesian inference are presented. 
\end{abstract}

\section{Introduction}
State-space models \citep{cappe2005inference} are widely applied in many fields, such as biology, chemistry, ecology, etc. 
Let us now introduce some notation. Consider an observable, discrete-time stochastic process $\{\bY_t\}_{t\geq t_0}$, $\bY_t\in\mathsf{Y}\subseteq \mathbb{R}^{d_y}$ and a latent and unobserved continuous-time stochastic process $\{\bX_t\}_{t\geq t_0}$, $\bX_t\in\mathsf{X}\subseteq \mathbb{R}^{d_x}$.
Process $\bX_t\sim p(\bx_t|\bx_{s},\btheta_x)$ is assumed Markov with transition densities $p(\cdot)$, $s<t$. Processes $\{\bX_t\}$ and $\{\bY_t\}$ depend on their own (unknown) vector-parameters $\btheta_x$ and $\btheta_y$, respectively.
We consider $\{\bY_t\}$ as a measurement-error-corrupted version of $\{\bX_t\}$ and assume that observations for $\{\bY_t\}$ are conditionally independent given $\{\bX_t\}$.
The state-space model can be summarised as
\begin{equation}
\begin{cases}
\bY_t\sim f(\by_t|\bX_t,\btheta_y),\\
\bX_t\sim p(\bx_t|\bx_{s},\btheta_x), \qquad t_0\leq s<t, \quad \bX_0\sim p(\bx_0)
\end{cases}
\label{eq:state-space-general}
\end{equation}
where $\bX_0\equiv \bX_{t_0}$.
We assume $f(\cdot)$ a known density (or probability mass) function set by the modeller. 
Regarding the transition density $p(\bx_t|\bx_{s},\cdot)$, this is typically unknown except for very simple toy models.

Goal of our work is to estimate the parameters $(\btheta_x,\btheta_y)$ by maximum likelihood, using observations $\bY_{1:n}=(\bY_1,...,\bY_n)$ collected at discrete times $\{t_1,...,t_n\}$. Here $\bY_j\equiv \bY_{t_j}$ and we use $\bz_{1:n}$ to denote a generic sequence $(\bz_1,...,\bz_n)$. For ease of notation we refer to the vector $\btheta:=(\btheta_x,\btheta_y)$ as the object of our inference. 

Parameters inference for state-space models has been widely developed, and sequential Monte Carlo (SMC) methods are now considered the state-of-art when dealing with nonlinear/non-Gaussian state space models (see \citealp{kantas2015particle} for a review). Methodological advancements have especially considered Bayesian approaches. In Bayesian inference the goal is to derive analytically the posterior distribution $\pi(\btheta|\bY_{1:n})$ or, most frequently, implement an algorithm for sampling draws from the posterior. Sampling procedures are often carried out using Markov chain Monte Carlo (MCMC) or SMC embedded in MCMC procedures, see \cite{andrieu2009pseudo} and \cite{andrieu2010particle}. 

In this work we instead aim at maximum likelihood estimation of $\btheta$. Several methods for maximum likelihood inference in state-space models have been proposed in the literature, including the well-known EM algorithm \citep{Dempster1977}. The EM algorithm computes the conditional expectation of the complete-likelihood for the pair $(\{\bY_t\},\{\bX_t\})$ and then produces a (local) maximizer for the likelihood function based on the actual observations $\bY_{1:n}$. One of the difficulties is how to compute the conditional expectation of the state $\{\bX_t\}$ given the observations  $\bY_{1:n}$. This conditional expectation can be computed exactly with the Kalman filter when the state-space is linear and Gaussian \citep{cappe2005inference}, otherwise it has to be approximated. In this work we focus on a stochastic approximation of $\{\bX_t\}$. Therefore, we resort to  
a stochastic version of the EM algorithm, namely the Stochastic Approximation EM (SAEM) \citep{Delyon1999}. The problem is to generate, conditionally on the current value of $\btheta$ during the EM maximization, an appropriate ``proposal'' for the state $\{\bX_t\}$, and we use SMC to obtain such proposal. SMC algorithms \citep{doucet2001sequential} have already been coupled to stochastic EM algorithms (see e.g. \cite{HuysPaninski2006, HuysPaninski2009, Lindsten2013, Ditlevsen2014} and references therein). The simplest and most popular SMC algorithm, the bootstrap filter \citep{gordon1993novel}, is easy to implement and very general, explicit knowledge of the density $f(\by_t|\bX_t,\cdot)$ being the only requirement. Therefore the bootstrap filter is often a go-to option for practitioners. Alternatively, in order to select a path $\{\bX_t\}$ to feed SAEM with, in this paper we follow an approach based on approximate Bayesian computation (ABC) and specifically we use the ABC-SMC method for state-estimation proposed in \cite{jasra2012filtering}. We do not merely consider the algorithm by Jasra et al. within SAEM, but show in detail and discuss how SAEM-ABC-SMC (shortly SAEM-ABC) can in some cases outperform SAEM coupled with the bootstrap filter.
 
 We illustrate our SAEM-ABC approach for (approximate) maximum likelihood estimation of $\btheta$ using two case studies, a nonlinear Gaussian state-space model and a more complex state-space model based on stochastic differential equations. We also compare our method with the iterated filtering for maximum likelihood estimation \citep{ionides2015inference}, Gibbs sampling and particle marginal methods for Bayesian inference \citep{andrieu2009pseudo} and will also use a special SMC proposal function for the specific case of stochastic differential equations \citep{golightly2011bayesian}. 

The paper is structured as follows: in section \ref{sec:complete-likelihood} we introduce the standard SAEM algorithm and basic notions of ABC. In section \ref{sec:abc-smc} we propose a new method for maximum likelihood estimation by integrating an ABC-SMC algorithm within SAEM. Section \ref{sec:simulations} shows simulation results and section  \ref{sec:summary} summarize conclusions. An appendix includes technical details pertaining the simulation studies. Software code can be found online either at \url{https://github.com/umbertopicchini/SAEM-ABC} or as supplementary material in the version of this paper published on \textit{Computational Statistics} (\texttt{doi:10.1007/s00180-017-0770-y})\footnote{The version of the code available on \textit{Computational Statistics} is the one submitted at the time of the article acceptance. The one on GitHub might contain more recent amendments (if any).\label{footnote:code}}.
 
\section{The complete likelihood and stochastic approximation EM}
\label{sec:complete-likelihood}

Recall that $\bY_{1:n}=(\bY_1,...,\bY_n)$ denotes the available data collected at times $(t_1,...,t_n)$ and denote with $\bX_{1:n}=(\bX_1,...,\bX_n)$ the corresponding unobserved states. We additionally set $\bX_{0:n}=(\bX_0,\bX_{1:n})$ for the vector including an initial (fixed or random) state $\bX_0$, that is $\bX_1$ is generated as $\bX_1\sim p(\bx_1|\bx_0)$. When the transition densities $p(\bx_j|\bx_{j-1})$ are available in closed form ($j=1,...,n$), the likelihood function for $\btheta$ can be written as (here we have assumed a random initial state with density $p(\bX_0)$)
\begin{align}
p(\bY_{1:n};\btheta) &= \int p_{\bY,\bX}(\bY_{1:n},\bX_{0:n}\, ; \btheta)\,d\bX_{0:n} = \int p_{\bY|\bX}(\bY_{1:n}|\bX_{0:n}\, ; \btheta)p_{\bX}(\bX_{0:n}; \btheta)\,d\bX_{0:n}\nonumber\\
&=\int p(\bX_0)\biggl\{\prod_{j=1}^n f(\bY_j|\bX_j;\btheta)p(\bX_j|\bX_{j-1};\btheta)\biggr\}d\bX_0\cdots d\bX_n  \label{eq:likelihood}
\end{align}
where $p_{\bY,\bX}$ is the ``complete data likelihood'', $p_{\bY|\bX}$ is the conditional law of $\bY$ given $\bX$,  $p_{\bX}$ is the law of $\bX$, $f(\bY_j|\bX_j;\cdot)$ is the conditional density of $\bY_j$  as in \eqref{eq:state-space-general} and $p_{\bX}(\bX_{0:n}; \cdot)$ the joint density of $\bX_{0:n}$. The last equality in \eqref{eq:likelihood} exploits the notion of conditional independence of observations given latent states and the Markovian property of $\{\bX_t\}$.  
In general the likelihood \eqref{eq:likelihood} is not explicitly known either because the integral is multidimensional and because expressions for transition densities are typically not available except for trivial toy models. 

In addition, when an exact simulator for the solution of the dynamical process associated with the Markov process $\{\bX_t\}$ is unavailable, hence it is not possible to sample from $p(\bX_j|\bX_{j-1};\btheta)$, numerical discretization methods are required. Without loss of generality, say that we have equispaced sampling times such that $t_j=t_{j-1}+\Delta$, with $\Delta>0$. Now introduce a discretization for the interval $[t_1,t_n]$ given by $\{\tau_1,\tau_h,...,\tau_{Rh},...,\tau_{nRh}\}$ where $h=\Delta/R$ and $R\geq 1$. We take $\tau_1=t_1$, $\tau_{nRh}=t_n$ and therefore $\tau_{i}\in \{t_1,....,t_n\}$ for $i=1,Rh,2Rh,...,nRh$. We denote with $N$ the number of elements in the discretisation $\{\tau_1,\tau_h,...,\tau_{Rh},...,\tau_{nRh}\}$ and with $\bX_{1:N}= (\bX_{\tau_1}, \ldots, \bX_{\tau_N})$ the corresponding values of $\{\bX_t\}$ obtained when using a given numerical/approximated method of choice. Then the likelihood function becomes
\begin{align*}
p(\bY_{1:n};\btheta) &= \int p_{\bY,\bX}(\bY_{1:n},\bX_{0:N}\, ; \btheta)\,d\bX_{0:N} = \int p_{\bY|\bX}(\bY_{1:n}|\bX_{0:N}\, ; \btheta)p_{\bX}(\bX_{0:N}; \btheta)\,d\bX_{0:N}\\
&=\int \biggl\{\prod_{j=1}^n f(\bY_j|\bX_j;\btheta)\biggr\}p(\bX_0)\prod_{i=1}^N p(\bX_i|\bX_{i-1};\btheta)d\bX_0\cdots d\bX_N, 
\end{align*}
where the product in $j$ is over the $\bX_{t_j}$ and the product in $i$ is over the $\bX_{\tau_i}$.

\subsection{The standard SAEM algorithm}\label{sec:standard-SAEM}
The EM algorithm introduced by  \cite{Dempster1977} is a classical approach to estimate parameters by maximum likelihood for models with non-observed or incomplete data. Let us briefly cover the EM principle. 
 The complete data of the model is  $(\bY_{1:n},\bX_{0:N})$, where $\bX_{0:N}\equiv \bX_{0:n}$ if numerical discretization is not required, and for ease of writing we denote this as $(\bY,\bX)$ in the remaining of this section.
The EM algorithm maximizes the function
 $Q(\btheta|\btheta')=\mathbb{E}(L_c(\bY,\bX;\btheta)|\bY;\btheta')$ 
in two steps, where $L_c(\bY,\bX;\btheta):=\log p_{\bY,\bX}(\bY,\bX;\btheta)$ is the log-likelihood of the complete data and $\mathbb{E}$ is the conditional expectation under the conditional distribution $p_{\bX|\bY} (\cdot ; \btheta')$. 
At the $k$-th iteration of a maximum (user defined) number of evaluations $K$, the E-step is the
evaluation of  $Q_k(\btheta)=Q(\btheta \, \vert \,\hat{\btheta}^{(k-1)})$, whereas the M-step
updates $\hat{\btheta}^{(k-1)}$ by maximizing $Q_{k}(\btheta)$.
For cases in which the E-step has no analytic form,   
\cite{Delyon1999}
introduce a stochastic version of the EM algorithm (SAEM)  which
evaluates the integral $Q_k(\btheta)$ by a stochastic approximation procedure.
The authors prove the convergence of this algorithm under general
conditions if $L_c(\bY,\bX;\btheta)$ belongs to the regular  exponential family
\[
L_c(\bY,\bX;\btheta)= -\Lambda(\btheta)+\langle \bS_c(\bY,\bX),\Gamma(\btheta)\rangle,
\]
where $\left\langle .,.\right\rangle$ is the scalar product, $\Lambda$ and $\Gamma$ are two functions of $\btheta$ and  $\bS_c(\bY,\bX)$ is  the minimal sufficient statistic of the complete model.  The E-step is then divided into a
simulation step (S-step) of the missing data $\bX^{(k)}$ under the
conditional distribution $p_{\bX|\bY}(\cdot;\hat{\btheta}^{(k-1)})$ and a stochastic
approximation step (SA-step) of the conditional expectation, using $(\gamma_k)_{k\geq 1}$ a sequence of real numbers in $[0,1]$, such that $\sum_{k=1}^\infty\gamma_k=\infty$ and $\sum_{k=1}^\infty\gamma_k^2<\infty$. This SA-step approximates  $\mathbb{E}\left\lbrack  \bS_c(\bY,\bX) \vert  \hat{\btheta}^{(k-1)} \right\rbrack$ at each iteration by the value $\bs_k$ defined recursively as follows
\begin{equation}
\bs_{k}=\bs_{k-1}+\gamma_k( \bS_c(\bY,\bX^{(k)})-\bs_{k-1}).\label{eq:summaries-simulation}
\end{equation}
The M-step is thus the update of the estimates $\hat{\btheta}^{(k-1)}$
\begin{equation}
\hat{\btheta}^{(k)}= \arg \max_{\btheta \in \Theta} \left(-\Lambda(\btheta)+\langle \bs_{k},\Gamma(\btheta)\rangle \right).\label{eq:m-step}
\end{equation}
The starting $\bs_0$ can be set to be a vector of zeros. The procedure above can be carried out iteratively for $K$ iterations. 
The proof of the convergence of SAEM requires the sequence $\gamma_k$   to be such that $\sum_{k=1}^\infty \gamma_k=\infty$ and $\sum_{k=1}^\infty \gamma_k^2<\infty$. A typical choice is to consider a warmup period with   $\gamma_k=1$ for the first $K_1$ iterations and then $\gamma_k=(k-K_1)^{-1}$ for $k\geq K_1$ (with $K_1<K$). 
Parameter $K_1$ has to be chosen by the user. However inference results are typically not very sensitive to this tuning parameter. Typical values are $K_1=250$ or $300$ and $K=400$, see for example  \cite{lavielle2014mixed}. 
Usually, the simulation step of the hidden trajectory $\bX^{(k)}$ conditionally on the observations $\bY$ cannot be directly performed. \cite{Lindsten2013} and  \cite{Ditlevsen2014} use a sequential Monte Carlo algorithm to perform the simulation step for state-space models.  We propose to resort to approximate Bayesian computation (ABC) for this simulation step. 

It can be noted that the generation of $\bs_k$ in \eqref{eq:summaries-simulation} followed by corresponding parameter estimates $\hat{\btheta}^{(k)}$ in \eqref{eq:m-step} is akin to two steps of a Gibbs sampling algorithm, except that here $\hat{\btheta}^{(k)}$ is produced by a deterministic step (for given $\bs_k$). We comment further on this aspect in section \ref{sec:gibbs}.

\subsection{The SAEM algorithm coupled to an ABC simulation step}
Approximate Bayesian Computation (ABC, see \citealp{marin-et-al(2011)} for a review) is a class of probabilistic algorithms allowing sampling from an approximation of a posterior distribution. The most typical use of ABC is when posterior inference for $\btheta$ is the goal of the analysis and the purpose is to sample draws from the approximate posterior $\pi_{\delta}(\btheta|\bY)$. Here and in the following $\bY\equiv\bY_{1:n}$. The parameter $\delta>0$ is a threshold influencing the quality of the inference, the smaller the $\delta$ the more accurate the inference, and  $\pi_{\delta}(\btheta|\bY)\equiv \pi(\btheta|\bY)$ when $\delta=0$. However in our study we are not interested in conducting Bayesian inference for $\btheta$. We will use ABC to sample from an approximation to the posterior distribution $\pi(\bX_{0:N}|\bY;\btheta)\equiv p(\bX_{0:N}|\bY;\btheta)$, that is for a fixed value of $\btheta$, we wish to sample from $\pi_{\delta}(\bX_{0:N}|\bY;\btheta)$ (recall from section \ref{sec:standard-SAEM} that when feasible we can take $N\equiv n$). 
For simplicity of notation, in the following we avoid specifying the dependence on the current value of $\btheta$, which has to be assumed as a deterministic unknown. There are several ways to generate a ``candidate'' $\bX^*_{0:N}$: for example we might consider ``blind'' forward simulation, meaning that $\bX^*_{0:N}$ is simulated from $p_{\bX}(\bX_{0:N})$ and therefore unconditionally on data (i.e. the simulator is blind with respect to data). Then $\bX^*_{0:N}$ is accepted if the corresponding $\bY^*$ simulated from $f(\cdot|\bX^*_{1:n})$ is ``close'' to $\bY$, according to the threshold $\delta$, where $\bX^*_{1:n}$ contains the interpolated values of $\bX^*_{0:N}$ at sampling times $\{t_1,...,t_n\}$ and $\bY^*\equiv \bY^*_{1:n}$. Notice that the appeal of the methodology is that  knowledge of the probabilistic features of the data generating model is not necessary, meaning that even if the transition densities $p(\bX_j|\bX_{j-1})$ are not known (hence $p_{\bX}$ is unknown) it is enough to be able to simulate from the model (using a numerical scheme if necessary) hence draws $\bX^*_{0:N}$ are produced by forward-simulation regardless the explicit knowledge of the underlying densities.
  
Algorithm \ref{alg:accept-reject} illustrates a generic iteration $k$ of a SAEM-ABC method, where the current value of the parameters is $\hat{\btheta}^{(k-1)}$ and an updated value of the estimates is produced as $\hat{\btheta}^{(k)}$.
\begin{algorithm}
\caption{A generic iteration of SAEM-ABC using acceptance-rejection}\label{alg:accept-reject}
\begin{algorithmic}
\State {\bf Simulation step:} here we update $\bX^{(k)}$  using an ABC procedure sampling from $\pi_{\delta}(\bX|\bY; \hat\btheta^{(k-1)})$:
	\begin{itemize}
	\item[] {\bf Repeat}
	\begin{itemize}
	\item Generate a candidate $\bX^*$ from the latent model dynamics conditionally on $\hat\btheta^{(k-1)}$, either by numerical methods or using the transition density (if available) i.e. by generating using the exact law $p_{\bX}(\cdot;  \hat\btheta^{(k-1)})$.
	\item Generate $\bY^*$ from the error model $f(\bY^*|\bX^*)$.
	\end{itemize}
	\item[] {\bf Until $\rho(\bY^*,\bY)\leq \delta$}
	\end{itemize}
	\item[] Set $\bX^{(k)}= \bX^*$
\item {\bf Stochastic Approximation step :} update of the sufficient statistics
	$$\bs_{k}  =  \bs_{k-1} + \gamma_k\,\left(\bS_c(\bY, \bX^{(k)}) -\bs_{k-1}\right)$$
\item{\bf Maximisation step:} update $\btheta$
	$$\hat{\btheta}^{(k)}= \arg \max_{\theta \in \Theta} \left(-\Lambda(\btheta)+\langle \bs_{k},\Gamma(\btheta)\rangle \right)$$

\end{algorithmic}
\end{algorithm}
By iterating the procedure $K$ times, the resulting $\hat{\btheta}^{(k)}$ is the maximizer for an approximate likelihood (the approximation implied by using ABC). The ``repeat loop'' can be considerably expensive using a distance $\rho(\bY^*,\bY)\leq \delta$ as acceptance of $\bY^*$ (hence acceptance of $\bX^*$) is a rare event for $\delta$ reasonably small. If informative low-dimensional statistics $\eta(\cdot)$ are available, it is recommended to consider $\rho(\eta(\bY^*),\eta(\bY))$ instead.

However algorithm \ref{alg:accept-reject} is not appropriate for state-space models, because the entire candidate trajectory $\bX^*$ is simulated blindly to data (an alternative approach is considered in section \ref{sec:abc-smc}).  If we consider for a moment $\bX^*\equiv \bX^*_{0:N}$ as a generic unknown, ideally we would like to sample from the posterior $\pi(\bX^*|\bY)$, which is proportional to
\begin{equation}
\pi(\bX^*|\bY)\propto f(\bY|\bX^*)\pi(\bX^*)\label{eq:filtering-density}
\end{equation}
for a given ``prior'' distribution $\pi(\bX^*)$. For some models sampling from such posterior is not trivial, for example when $\bX$ is a stochastic process and in that case sequential Monte Carlo methods can be used as described in section \ref{sec:abc-smc}. A further layer of approximation is introduced when $\pi(\bX^*|\bY)$ is analytically ``intractable'', and specifically when $f(\bY|\bX)$ is unavailable in closed form (though we always assume $f(\cdot|\cdot)$ to be known and that it is possible to evaluate it pointwise) or is computationally difficult to approximate but it is easy to sample from. In this case ABC methodology turns useful, and an ABC approximation to $\pi(\bX^*|\bY)$ can be written as
\begin{equation}
\pi_{\delta}(\bX^*|\bY)\propto \int J_{\delta}(\bY,\bY^*) \underbrace{f(\bY^*|\bX^*)\pi(\bX^*)}_{\propto \pi(\bX^*|\bY^*)}d\bY^*. \label{eq:abc-posterior}
\end{equation}
Here $J_{\delta}(\cdot)$ is some function that depends on $\delta$ and weights the intractable posterior based on simulated data $\pi(\bX^*|\bY^*)$  with high values in regions where $\bY$ and $\bY^*$ are similar; therefore we would like (i) $J_{\delta}(\cdot)$ to give higher rewards to proposals $\bX^*$ corresponding to $\bY^*$ having values close to $\bY$. In addition (ii) $J_{\delta}(\bY,\bY^*)$ is assumed to be a constant when $\bY^*=\bY$ (i.e. when $\delta=0$) so that $J_{\delta}$ is absorbed into the proportionality constant and the \textit{exact} marginal posterior $\pi(\bX^*|\bY)$ is recovered. Basically the use of \eqref{eq:abc-posterior} means to simulate $\bX^*$ from its prior (the product of transition densities), then plug such draw into $f(\cdot|\bX^*)$ to simulate $\bY^*$, so that $\bX^*$ will be weighted by $J_{\delta}(\bY,\bY^*)$.
A common choice for $J_{\delta}(\cdot)$ is the uniform kernel
\[J_{\delta}(\bY,\bY^*)\propto \mathbb{I}_{\{\rho(\bY^*,\bY)\leq\delta\}}\]
where $\rho(\bY,\bY^*)$ is some measure of closeness between $\bY^*$ and $\bY$ and $\mathbb{I}$ is the indicator function. A further popular possibility is the Gaussian kernel
\begin{equation}\label{eq:GaussianKernel}
J_{\delta}(\bY, \bY^{*})\propto e^{-(\bY^{*}-\bY)'(\bY^{*}-\bY)/2\delta^2}
\end{equation}
where $'$ denotes transposition, so that $J_{\delta}(\bY, \bY^{*})$ gets larger when $\bY^{*}\approx \bY$. 

However one of the difficulties is that, in practice, $\delta$ has to be set as a compromise between statistical accuracy (a small positive $\delta$) and computational feasibility ($\delta$ not too small). Notice that a proposal's acceptance can be significantly enhanced when the posterior \eqref{eq:abc-posterior} is conditional on summary statistics of data $\eta(\bY)$, rather than $\bY$ itself, and in such case we would consider $\rho(\eta(\bY),\eta(\bY^*))$. However, in practice for dynamical models it is difficult to identify ``informative enough'' summary statistics $\eta(\cdot)$, but see \cite{martin2014approximate} and \cite{picchini2015accelerating}. Another important problem with the strategy outlined above is that ``blind simulation'' for the generation of the entire time series $\bX^*$ is often poor. 
 In fact, even when the current value of $\btheta$ is close to its true value, proposed trajectories rarely follow measurements when (a) the dataset is a long time series and/or (b) the model is highly erratic, for example when latent dynamics are expressed by a stochastic differential equation (section \ref{sec:theoph}). For these reasons, sequential Monte Carlo (SMC) \citep{Cappe2007} methods have emerged as the most  successful solution for filtering in non-linear non-Gaussian state-space models.

Below we consider the ABC-SMC methodology from \cite{jasra2012filtering}, which proves effective for state-space models.

\section{SAEM coupled with an ABC-SMC algorithm for filtering}\label{sec:abc-smc}

Here we consider a strategy for filtering that is based on an ABC version of sequential Monte Carlo sampling, as presented in \cite{jasra2012filtering}, with some minor modifications. The advantage of the methodology is that the generation of proposed trajectories is sequential, that is the ABC distance is evaluated ``locally'' for each observational time point. In fact, what is evaluated is the proximity of trajectories/particles to each data point $\bY_j$, and ``bad trajectories'' are killed thus preventing the propagation of unlikely states to the next observation $\bY_{j+1}$ and so on. For simplicity we consider the case $N\equiv n$, $h\equiv \Delta$. The algorithm samples from the following target density at time $t_{n'}$ ($n'\leq n$):
$$
\pi_{n',\delta}(\bX_{1:n'},\bY_{1:n'}^*|\bY_{1:n'})\propto \prod_{j=1}^{n'}J_{j,\delta}(\bY_j,\bY_j^*)f(\bY_j^*|\bX_j)p(\bX_{j}|\bX_{j-1})
$$
and for example we could take $J_{j,\delta}(\by_j,\by_j^*)=\mathbb{I}_{A_{\delta,\by_j}}(\by_j^*)$ with $A_{\delta,\by_j}=\{\by^*_j;$  $\rho(\eta(\by^*_j),\eta(\by_j))<\delta\}$ as in \cite{jasra2012filtering} or the Gaussian kernel (\ref{eq:GaussianKernel}). 

The ABC-SMC procedure is set in algorithm \ref{alg:abc-smc} with the purpose to propagate forward $M$ simulated states (``particles''). After algorithm \ref{alg:abc-smc} is executed, we select a single trajectory by retrospectively looking at the genealogy of the generated particles, as explained further below.
\begin{algorithm}
\caption{ABC-SMC for filtering}\label{alg:abc-smc}
\begin{algorithmic}
\State Step 0. Set $j=1$. For $m=1,...,M$ sample $\bX_1^{(m)}\sim p(\bX_0)$, $\bY_1^{*(m)}\sim f(\cdot|\bX_1^{(m)})$, compute weights $W_1^{(m)}=J_{1,\delta}(\bY_1,\bY_1^{*(m)})$ and normalize weights $w_1^{(m)}:=W_1^{(m)}/\sum_{m=1}^M W_1^{(m)}$.
\State Step 1.
   \If{$ESS(\{w_j^{(m)}\})<\bar{M}$} 
\State resample $M$   particles $\{\bX_j^{(m)},w_j^{(m)}\}$ and set $W_j^{(m)}=1/M$. 
   \EndIf
\State Set $j:=j+1$ and if $j=n+1$, stop.
\State Step 2. For $m=1,...,M$ sample $\bX_j^{(m)}\sim p(\cdot|\bX_{j-1}^{(m)})$ and $\bY_j^{*(m)}\sim f(\cdot|\bX_j^{(m)})$. Compute 
$$W_j^{(m)}:=w_{j-1}^{(m)}J_{j,\delta}(\bY_j,\bY_j^{*(m)})$$
normalize weights $w_j^{(m)}:=W_{j}^{(m)}/\sum_{m=1}^M W_j^{(m)}$ and go to step 1.
\end{algorithmic}
\end{algorithm}
The quantity ESS is the effective sample size (e.g. \citealp{liu2008monte}) often estimated as $ESS(\{w_j^{(m)}\})=1/\sum_{m=1}^M(w_j^{(m)})^2$ and taking values between 1 and $M$. When considering an indicator function for $J_{j,\delta}$, the ESS coincides with the number of particles having positive weight \citep{jasra2012filtering}. Under such choice the integer $\bar{M}\leq M$ is a lower bound (threshold set by the experimenter) on the number of particles with non-zero weight. However in our experiments we use a Gaussian kernel and since in the examples in section \ref{sec:simulations} we have a scalar $Y_j$, the kernel is defined as
\begin{equation}
J_{j,\delta}(Y_j,Y_j^{*(m)})=\frac{1}{\delta} e^{-(Y_j^{*(m)}-Y_j)^2/(2\delta^2)}\label{eq:gauss-kernel}
\end{equation}
so that weights $W_j^{(m)}$ are larger for particles having $Y_j^{*(m)}\approx Y_j$.  
We consider ``stratified resampling'' \citep{kitagawa1996monte} in step 1 of algorithm \ref{alg:abc-smc}. 

In addition to the procedure outlined in algorithm \ref{alg:abc-smc}, once the set of weights $\{w_n^{(1)},...,w_n^{(M)}\}$ is available at time $t_n$, we propose to follow \cite{andrieu2010particle} (see their PMMH algorithm) and sample a single index from the set $\{1,...,M\}$ having associated probabilities $\{w_n^{(1)},...,w_n^{(M)}\}$. Denote with $m'$ such index and with $a_j^m$ the ``ancestor'' of the generic $m$th particle sampled at time $t_{j+1}$, with $1\leq a_j^m \leq M$ ($m=1,...,M$, $j=1,...,n$). Then we have that particle $m'$ has ancestor $a_{n-1}^{m'}$ and in general particle $m''$ at time $t_{j+1}$ has ancestor $b_j^{m''}:=a_j^{b^{m''}_{j+1}}$, with $b_n^{m'}:=m'$. Hence, at the end of algorithm \ref{alg:abc-smc} we can sample $m'$ and construct its genealogy: the sequence of states $\{X_t\}$ resulting from the genealogy of $m'$ is the chosen path that will be passed to SAEM, see algorithm 
\ref{alg:saem-abc-smc}.  The selection of this path is crucially affected by ``particles impoverishment'' issues, see below. [An alternative procedure, which we do not pursue here, is to sample at time $t_n$ not just a single index $m'$, but instead sample with replacement say $G\geq 1$ times from the set $\{1,...,M\}$ having associated probabilities $\{w_n^{(1)},...,w_n^{(M)}\}$. Then construct the genealogy for each of the $G$ sampled indeces, and for each of the resulting $G$ sampled paths $\bX^{g,k}$ calculate the corresponding vector-summaries $\bS^{g,k}_c:=\bS_c(\bY,\bX^{g,k})$, $g=1,...,G$. Then it would be possible to take the sample average of those $G$ summaries $\bar{\bS}^{k}_c$ as in \cite{kuhn2005maximum}, and plug this average in place of $\bS_c(\bY,\bX^{(k)})$ in step 3 of algorithm \ref{alg:saem-abc-smc}. Clearly this approach increases the computational complexity linearly with $G$.]

Notice that in \cite{jasra2012filtering} $n$ ABC thresholds $\{\delta_1,...,\delta_n\}$ are constructed, one threshold for each corresponding sampling time in $\{t_1,...,t_n\}$: these thresholds do not need to be set by the user but can be updated adaptively using a stochastic data-driven procedure. This is possible because the ABC-SMC algorithm in \cite{jasra2012filtering} is for filtering only, that is $\btheta$ is a fixed known quantity. However in our scenario $\btheta$ is unknown, and letting the thresholds vary adaptively (and randomly) between each pair of iterations of a parameter estimation algorithm is not appropriate. This is because the evaluation of the likelihood function at two different iterations $k'$ and $k''$ of SAEM would depend on  a procedure determining corresponding (stochastic) sequences $\{\delta_1^{(k')},...,\delta_n^{(k')}\}$ and $\{\delta_1^{(k'')},...,\delta_n^{(k'')}\}$. Therefore the likelihood maximization would be affected by the random realizations of the sequences of thresholds. In our case, we let the threshold vary \textit{deterministically}: that is we choose a $\delta$ common to all time points $\{t_1,...,t_n\}$ and execute a number of SAEM iterations using such threshold. Then we deterministically decrease the threshold according to a user-defined scheme and execute further SAEM iterations, and so on. Our SAEM-ABC procedure is detailed in algorithm \ref{alg:saem-abc-smc} with a user-defined sequence $\delta_1>\cdots >\delta_L>0$ where each $\delta_l$ is used for $k_l$ iterations, so that $k_1+\cdots+k_L=K$ ($l=1,...,L$). Here $K$ is the number of SAEM iterations, as defined at the end of section \ref{sec:standard-SAEM}. In our applications we show how the algorithm is not overly sensitive to small variations in  the $\delta$'s.

Regarding the choice of $\delta$ values in applications, recall that the interpretation of $\delta$ in equation \eqref{eq:gauss-kernel} is that of the standard deviation for a perturbed model. This implies that a synthetic observation at time $t_j$, denoted with $Y^*_j$, can be interpreted as a perturbed version of $Y_j$, where the observed $Y_j$ is assumed generated from the state-space model in equation (1), while $Y^*_j\sim \mathcal{N}(Y_j,\delta^2)$. With this fact in mind, $\delta$ can easily be chosen to represent some deviation from the actual observations. Therefore, typically it is enough to look at the time evolution of the data, to guess at least the order of magnitude of the starting value for $\delta$.

Finally, in our applications we compare SAEM-ABC with SAEM-SMC. SAEM-SMC is detailed in algorithm \ref{alg:saem-smc}: it is structurally the same as algorithm \ref{alg:saem-abc-smc} except that it uses the bootstrap filter to select the state trajectory. The bootstrap filter \citep{gordon1993novel} is just like algorithm \ref{alg:abc-smc}, except that no simulation from the observations equation is performed (i.e. the $\bY_j^{*(m)}$ are not generated) and $J_{j,\delta}(\bY_j,\bY_j^{*(m)})$ is replaced with $f(\bY_j|\bX_j^{(m)})$, hence there is no need to specify the $\delta$. The trajectory $\bX^{(k)}$ selected by the bootstrap filter at $k$th SAEM iteration is then used to update the statistics $\bs_k$. 

As studied in detail in section \ref{ex:theophylline-results}, the function $J_{t,\delta}$ has an important role in approximating the density $\pi(\bX_{1:t}|\bY_{1:t})$ at a generic time instant $t$. Indeed $\pi_\delta(\bX_{1:t}|\bY_{1:t})=\int \pi_{\delta}(\bX_{1:t},\bY_{1:t}^*|\bY_{1:t})d\bY^*_{1:t}$, the ABC approximation to $\pi(\bX_{1:t}|\bY_{1:t})$, is 
\[
\pi_\delta(\bX_{1:t}|\bY_{1:t})=
\sum_{m=1}^M W_t^{(m)}\xi_{\bx_{0:t}}(\bx_{0:t}^{(m)})
\]
with $\xi(\cdot)$ the Dirac measure and $W_t^{(m)}\propto J_{t,\delta}(\bY_t^{*(m)},\bY_t)$. This means that for a small $\delta$ (which is set by the user) $J_\delta$ assigns large weights only to very promising particles, that is those particles $\bX_t^{(m)}$ associated to a $\bY^{*(m)}_t$ very close to $\bY_t$. Instead, suppose that $f(\cdot|\bX_t,\btheta_y)\equiv \mathcal{N}(\cdot|\bX_t;\sigma^2_\varepsilon)$ where  $\mathcal{N}(\cdot|a;b)$ is a Gaussian distribution with mean $a$ and variance $b$, then the bootstrap filter underlying SAEM-SMC assigns weights proportionally to the measurements density $f(\bY_t|\bX_t;\sigma_\varepsilon)$, which is affected by the currently available (and possibly poor) value of $\sigma_\varepsilon$, when $\sigma_\varepsilon$ is one of the unknowns to be estimated. When $\sigma_\varepsilon$ is overestimated, as in section \ref{ex:theophylline-results}, there is a risk to sample particles which are not really ``important''. Since trajectories selected in step 2 of algorithm \ref{alg:saem-smc} are  drawn from $\pi(\bX_{1:t}|\bY_{1:t})$, clearly the issues just highlighted contribute to bias the inference.

\begin{algorithm}
\caption{SAEM-ABC using a particle filter}\label{alg:saem-abc-smc}
\begin{algorithmic}
\State Step 0. Set parameters starting values $\hat{\btheta}^{(0)}$, set $M$, $\bar{M}$ and $k:=1$. Set the sequence $\{\delta_1,...,\delta_L\}$ and $\delta:=\delta_1$.
\State Step 1. For fixed $\hat{\btheta}^{(k-1)}$ apply the ABC-SMC algorithm \ref{alg:abc-smc} with threshold $\delta$, $M$ particles and particles threshold $\bar{M}$. 
\State 2 Sample an index $m'$ from the probability distribution $\{w_n^{(1)},...,w_n^{(M)}\}$ on $\{1,...,M\}$ and form the path $\bX^{(k)}$ resulting from the genealogy of $m'$.
\State Step 3. {\bf Stochastic Approximation step :} update of the sufficient statistics
	$$\bs_{k}  =  \bs_{k-1} + \gamma_k\,\left(\bS_c(\bY, \bX^{(k)}) -\bs_{k-1}\right)$$
\State Step 4. {\bf Maximisation step:} update $\btheta$
	$$\hat{\btheta}^{(k)}= \arg \max_{\btheta \in \Theta} \left(-\Lambda(\btheta)+\langle \bs_{k},\Gamma(\btheta)\rangle \right)$$
Set $k:=k+1$. If $k\in \{k_1,...,k_L\}$, e.g. if $k=k_l$, set $\delta:=\delta_l$. Go to step 1.
\end{algorithmic}
\end{algorithm}

\begin{algorithm}
\caption{SAEM-SMC}\label{alg:saem-smc}
\begin{algorithmic}
\State Step 0. Set parameters starting values $\hat{\btheta}^{(0)}$, set $M$, $\bar{M}$ and $k:=1$. 
For fixed $\hat{\btheta}^{(k-1)}$ apply the bootstrap filter below using  $M$ particles and particles threshold $\bar{M}$.
\State  Step 1a. Set $j=1$. For $m=1,...,M$ and conditionally on $\hat{\btheta}^{(k-1)}$ sample $\bX_1^{(m)}\sim p(\bX_0)$,  compute weights $W_1^{(m)}=f(\bY_1|\bX_1^{(m)})$ and normalize weights $w_1^{(m)}:=W_1^{(m)}/\sum_{m=1}^M W_1^{(m)}$.
\State   Step 1.b. \If{$ESS(\{w_j^{(m)}\})<\bar{M}$} 
  resample $M$   particles $\{\bX_j^{(m)},w_j^{(m)}\}$ and set $W_j^{(m)}=1/M$. 
  \EndIf
\State  Set $j:=j+1$. If $j=n+1$ go to step 2. If $j\leq n$ go to step 1.c.
\State   Step 1.c. For $m=1,...,M$ and conditionally on $\hat{\btheta}^{(k-1)}$ sample $\bX_j^{(m)}\sim p(\cdot|\bX_{j-1}^{(m)})$. Compute 
$$W_j^{(m)}:=w_{j-1}^{(m)}f(\bY_j|\bX_j^{(m)})$$
   normalize weights $w_j^{(m)}:=W_{j}^{(m)}/\sum_{m=1}^M W_j^{(m)}$ and go to step 1.b.

\State Step 2.  Sample an index $m'$ from the probability distribution $\{w_n^{(1)},...,w_n^{(M)}\}$ on $\{1,...,M\}$ and form the path $\bX^{(k)}$ resulting from the genealogy of $m'$.
\State Step 3. {\bf Stochastic Approximation step :} update of the sufficient statistics
	$$\bs_{k}  =  \bs_{k-1} + \gamma_k\,\left(\bS_c(\bY, \bX^{(k)}) -\bs_{k-1}\right)$$
\State Step 4. {\bf Maximisation step:} update $\btheta$
	$$\hat{\btheta}^{(k)}= \arg \max_{\btheta \in \Theta} \left(-\Lambda(\btheta)+\langle \bs_{k},\Gamma(\btheta)\rangle \right)$$
Set $k:=k+1$. Go to step 1a.
\end{algorithmic}
\end{algorithm}

A further issue is studied in section \ref{sec:impoverishment}. There we explain how SAEM-ABC, despite being an approximate version of SAEM-SMC (due to the additional approximation induced by using a strictly positive $\delta$) can in practice outperform SAEM coupled with the simple bootstrap filter, because of ``particles impoverishment'' problems. Particles impoverishment is a pathology due to frequently implementing particles resampling: the resampling step reduces the ``variety'' of the particles, by duplicating the ones with larger weights and killing the others. We show that when particles fail to get close to the targeted observations, then resampling is frequently triggered, this degrading the variety of the particles. However with an ABC filter, particles receive some additional weighting, due to the function $J_{j,\delta}$ in equation \eqref{eq:gauss-kernel} taking values $J_{j,\delta}>1$ for small $\delta$ when $\bY^*_j\approx \bY_j$, which allows a larger number of particles to have a  non-negligible weight and therefore different particles are resampled, this increasing their variety (at least for the application in section \ref{sec:nonlingauss-largenoise}, but this is not true in general). This is especially relevant in early iterations of SAEM, where $\btheta^{(k)}$ might be far from its true value and therefore many particles might end far from data. By letting $\delta$ decrease not ``too fast'' as SAEM iterations increase, we allow many particles to contribute to the states propagation. However, as $\delta$ approaches a small value, only the most promising particles will contribute to selecting the path sampled in step 2 of algorithm \ref{alg:saem-abc-smc}.

These aspects are discussed in greater detail in sections \ref{sec:impoverishment} and \ref{ex:theophylline-results}. However, it is of course \textit{not true} that an ABC filter is in general expected to perform better than a non-ABC one. In the second example, section  \ref{sec:theoph}, an adapted (not ``blind to data'', i.e. conditional to the next observation) particle filter is used to treat the specific case of SDE models requiring numerical integration \citep{golightly2011bayesian}. The adapted filter clearly outperforms both SAEM-SMC and SAEM-ABC when the number of particles is very limited ($M=30$).

Notice, our SAEM-ABC strategy with a decreasing series of thresholds shares some similarity with tempering approaches. For example \cite{herbst2017} artificially inflate the observational noise variance pertaining $f(\by_t|\bX_t,\cdot)$, so that the particle weights have lower variance hence the resulting filter is more stable. More in detail, they construct a bootstrap filter where particles are propagated through a sequence of intermediate tempering steps, starting from  an observational distribution with  inflated  variance,  and 
then gradually reducing the variance to its nominal level.

\paragraph{Fisher Information matrix}	\label{sec:fisher}

The SAEM algorithm allows also to compute standard errors of the estimators, through the approximation of the Fisher Information matrix. This is detailed below, however notice that the algorithm itself advances between iterations without the need to compute such matrix (nor the gradient of the function to maximize). 
	The standard errors of the parameter estimates can be calculated
from the diagonal elements of the inverse of the Fisher information
matrix. Its direct evaluation 
is difficult because it has no explicit analytic form, however an estimate of the Fisher information
matrix can easily be implemented within SAEM as  proposed by \cite{Delyon1999} using the Louis' missing information principle \citep{louis1982finding}.
  
The Fisher information matrix $\ell(\btheta)=L(\bY; \btheta)$ can  be expressed as:
\begin{eqnarray*}
\partial^2_\theta  \ell(\btheta)& =& \mathbb{E}\left[\partial^2_\theta L_c(\bY,\bX;\btheta)|\bY,\btheta\right] \\
&+& \mathbb{E}\left[\partial_\theta L_c(\bY,\bX;\btheta)\;(\partial_\theta L_c(\bY,\bX;\btheta))'|\bY,\btheta\right] \\
&-& \mathbb{E}\left[\partial_\theta L_c(\bY,\bX;\btheta)|\bY,\btheta\right]\;\mathbb{E}\left[\partial_\theta L_c(\bY,\bX;\btheta)|\bY,\btheta\right]'
\end{eqnarray*}
where $'$ denotes transposition. An on-line estimation of the Fisher information is obtained using the stochastic approximation procedure of the SAEM algorithm as follows (see \citealp{lavielle2014mixed} for an off-line approach).
At the $(k+1)$th iteration of the algorithm, we evaluate the three following quantities:
\begin{eqnarray*}
\bG_{k+1}& =& \bG_k + \gamma_ {k} \left[\partial_\theta L_c(\bY,\bX^{(k)}, \btheta)
- \bG_k\right]\\
\bH_{k+1}& =& \bH_k + \gamma_ {k} \left[\partial^2_\theta L_c(\bY,\bX^{(k)}, \btheta) \right.\\
&&+\left. \partial_\theta L_c(\bY,\bX^{(k)}, \btheta)\;(\partial_\theta L_c(\bY,\bX^{(k)}, \btheta))'
- \bH_k\right]\\
\bF_{k+1} &=& \bH_{k+1}-\bG_{k+1}\;(\bG_{k+1})'.
\end{eqnarray*}
As the sequence $(\hat{\btheta}^{(k)})_k$ converges to the maximum of an approximate likelihood, the sequence $(\bF_k)_k$ converges to the corresponding approximate Fisher information matrix. It is possible to initialize $\bG_0$ and $\bH_0$ to be a vector and a matrix of zeros respectively. We stress that we do not make use of the (approximate) Fisher information during the optimization.

\section{Simulation studies}\label{sec:simulations}
Simulations were coded in MATLAB (except for R examples using the \texttt{pomp} package) and executed on a Intel Core i7-2600 CPU 3.40 GhZ. Software code can be found online either at \url{https://github.com/umbertopicchini/SAEM-ABC} or as supplementary material in the version of this paper published on \textit{Computational Statistics} (\texttt{doi:10.1007/s00180-017-0770-y}), see also the footnote on page \pageref{footnote:code}. For all examples we consider a Gaussian kernel for $J_{j,\delta}$ as in \eqref{eq:gauss-kernel}. As described at the end of section \ref{sec:standard-SAEM}, in SAEM we always set $\gamma_k=1$ for the first $K_1$ iterations and $\gamma_k=(k-K_1)^{-1}$ for $k\geq K_1$ as in \cite{lavielle2014mixed}.
All results involving ABC are produced using algorithm \ref{alg:saem-abc-smc} i.e. using trajectories selected via ABC-SMC. We compare our results with standard algorithms for Bayesian and ``classical'' inference, namely Gibbs sampling and particle marginal methods (PMM) \citep{andrieu2009pseudo} for Bayesian inference and the improved iterated filtering (denoted in literature as IF2) found in \cite{ionides2015inference} for maximum likelihood estimation. In order to perform a fair comparison between methods, we make use of well tested and maintained code to fit models with PMM and IF2 via the R \texttt{pomp} package \citep{king2015statistical}. 
All the methods mentioned above use sequential Monte Carlo algorithms (SMC), and their \texttt{pomp} implementation considers the bootstrap filter. We remark that our goal is not to consider specialized state-of-art SMC algorithms, with the notable exception mentioned below. Our focus is to compare the several inference methods above, while using the bootstrap filter for the trajectory proposal step: the bootstrap filter is also the approach considered in \cite{king2015statistical} and \cite{fasiolo2016comparison} hence it is easier for us to compare methods using available software packages such as \texttt{pomp}. However in section \ref{sec:theoph} we also use a particles sampler that conditions upon data and is suitable for state-space models having latent process expressed by a stochastic differential equation \citep{golightly2011bayesian}.

\subsection{Non-linear Gaussian state-space model}\label{sec:nonlingauss}

Consider the following Gaussian state-space model
\begin{align}
\begin{cases}
Y_j = X_j + \sigma_y\nu_j\\
X_j = 2\sin(e^{X_{j-1}})+\sigma_x\tau_j, \quad j=1,...,n
\end{cases}
\label{eq:nonlingauss-model}
\end{align}
with $\nu_j,\tau_j\sim N(0,1)$ i.i.d. and $X_0=0$. Parameters  $\sigma_x,\sigma_y>0$ are the only unknowns and therefore we conduct inference for $\btheta = (\sigma^2_x,\sigma^2_y)$.

We first construct the set of sufficient statistics corresponding to the complete log-likelihood $L_c(\bY,\bX)$. This is a very simple task since $Y_j|X_j\sim N(X_j,\sigma^2_y)$ and $X_j|X_{j-1}\sim N(2\sin(e^{X_{j-1}}),\sigma^2_x)$ and therefore it is easy to show that $S_{\sigma^2_x}=\sum_{j=1}^n (X_j-2\sin(e^{X_{j-1}}))^2$ and $S_{\sigma^2_y}=\sum_{j=1}^n (Y_j-X_{j})^2$ are sufficient for $\sigma^2_x$ and $\sigma^2_y$ respectively. By plugging these statistics into $L_c(\bY,\bX)$ and equating to zero the gradient of $L_c$ with respect to $(\sigma^2_x,\sigma^2_y)$, we find that the M-step of SAEM results in updated values for $\sigma^2_x$ and $\sigma^2_y$ given by $S_{\sigma^2_x}/n$ and $S_{\sigma^2_y}/n$ respectively. Expressions for the first, second and mixed derivatives, useful to obtain the Fisher information as in Section \ref{sec:fisher}, are given in appendix.

\subsubsection{Results}\label{sec:nonlingauss-largenoise}

\begin{figure}
\centering
\includegraphics[scale=0.35]{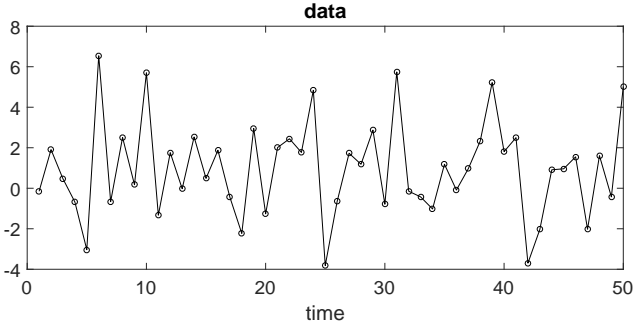}
\caption{\footnotesize{Nonlinear-Gaussian model: data when $\sigma_x=\sigma_y=2.23$.}}\label{fig:nonlingauss-data-largerror}
\end{figure}

\begin{figure}
\centering
\includegraphics[width=14cm,height=5cm]{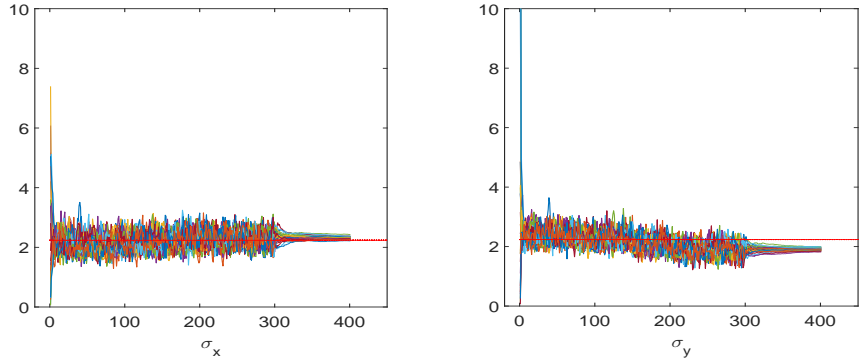}
\caption{\footnotesize{Non-linear Gaussian model: traces obtained with SAEM-ABC when using $M=1,000$ particles, $\bar{M}=200$ and $(\sigma_x,\sigma_y)=(2.23,2.23)$. Horizontal lines are the true parameter values.}}
\label{fig:nonlingauss-SAEM-ABC-iter}
\end{figure}

We generate $n=50$ observations for $\{Y_j\}$ with $\sigma^2_x=\sigma^2_y=5$, see Figure \ref{fig:nonlingauss-data-largerror}.
We first describe results obtained using SAEM-ABC. Since the parameters of interest are positive, for numerical convenience we work on the log-transformed versions $(\log\sigma_x,\log\sigma_y)$. Our setup consists in running 30 independent experiments with SAEM-ABC: for each experiment we simulate parameter starting values for $(\log\sigma_x,\log\sigma_y)$ independently generated from a bivariate Gaussian distribution with mean the true value of the parameters, i.e. $(\log\sqrt{5},\log\sqrt{5})$, and diagonal covariance matrix having values (2,2) on its diagonal. 
For all 30 simulations we use the same data and the same setup except that in each simulation we use different starting values for the parameters. 
For each of the 30 experiments we let the threshold $\delta$ decrease in the set of values $\delta\in\{2,1.7,1.3,1\}$ for a total of $K=400$ SAEM-ABC iterations, where we use $\delta=2$ for the first 80 iterations, $\delta=1.7$ for further 70 iterations, $\delta=1.3$ for further 50 iterations and $\delta=1$ for the remaining 200 iterations. The influence of this choice is studied below. As explained in section \ref{sec:abc-smc}, the largest value for $\delta$ can be set intuitively, by looking at Figure \ref{fig:nonlingauss-data-largerror}, where it is apparent that considering deviations $\delta=2$ of the simulated observations $Y_j^*$ from the actual observation $Y_j$ should be reasonable. For example the empirical standard deviation of the differences $|Y_{j}-Y_{j-1}|$ is about 2. Then we let $\delta$ decrease progressively as SAEM-ABC evolves.  We take $K_1=300$ as the number of SAEM warmup iterations and use different numbers of particles $M$ in our simulation studies, see Table \ref{tab:nonlingauss}. We impose resampling when the effective sample size ESS gets smaller than $\bar{M}$, for any attempted value of $M$. At first we show that taking $\bar{M}=200$ gives good results for SAEM-ABC but not for SAEM-SMC, see Table \ref{tab:nonlingauss}. Table \ref{tab:nonlingauss} reports the median of the 30 estimates and their $1^{st}-3^{rd}$ quartiles: we notice that $M=1,000$ particles are able to return satisfactory estimates when using SAEM-ABC. Figure \ref{fig:nonlingauss-SAEM-ABC-iter} shows the rapid convergence of the algorithm towards the true parameter values for all the 30 repetitions (though difficult to notice visually, several simulations start at locations very far from the true parameter values). Notice that it only required about 150 seconds to perform all 30 simulations: this is the useful return out of the effort of constructing the analytic quantities necessary to run SAEM. Also, the algorithm is not very sensitive to the choice of $\delta's$. For example, we also experimented with $\delta \in \{4, 3, 2, 1\}$ and we obtained very similar results, for example our thirty experiments with $(M,\bar{M})=(1000,200)$ resulted in medians (1st-3rd quartiles) $\hat{\sigma}_x=2.30$ [2.27,2.35], $\hat{\sigma}_y=1.90$ [1.86,1.91], which are very close to the ones in Table \ref{tab:nonlingauss}. Finally, notice that results are not overly sensitive to the way $\delta$ is decreased: for example, if we let $\delta$ decrease uniformly with steps of size 1/3, that is $\delta\in\{2, 1.67, 1.33, 1\}$ with $\delta=2$ for the first 50 iterations, then let it decrease every 50 iterations until $\delta=1$, we obtain (when $M=1,000$ and $\bar{M}=200$) $\hat{\sigma}_x=2.30$ [2.27,2.33], $\hat{\sigma}_y=1.91$ [1.87,1.95], compare with Table \ref{tab:nonlingauss}.

\begin{table}
\small
\centering
\begin{tabular}{lcccc}
\hline\\
  & $(M,\bar{M})$=(500,200) & $(M,\bar{M})$=(1000,200) & $(M,\bar{M})$=(2000,200) & $(M,\bar{M})$=(1000,20)   \\
\hline 
 & \multicolumn{4}{c}{$\sigma_x$ estimates (true = $2.23$)} \\
SAEM-ABC & 2.42 [2.39,2.47] & 2.30 [2.27,2.32] & 2.19 [2.17,2.24] & 1.88 [1.81,1.93]\\
SAEM-SMC & 2.54 [2.53,2.54] & 2.55 [2.54,2.56] & 2.55 [2.54,2.56] & 1.99 [1.85,2.14]\\
IF2* & 1.26 [1.21,1.41] & 1.35 [1.28,1.41] & 1.33 [1.28,1.40] & 1.35 [1.28,1.41] \\
\hline 
 &  \multicolumn{4}{c}{$\sigma_y$ estimates (true = $2.23$)} \\
SAEM-ABC & 1.90 [1.87,1.94] & 1.91 [1.88,1.95] & 1.87 [1.84,1.93] & 1.91 [1.89,1.96]\\
SAEM-SMC & 0.11 [0.10,0.13] & 0.06 [0.06,0.07] & 0.04 [0.03,0.04] & 1.23 [0.99,1.39]\\
IF2* & 1.62 [1.56,1.75] & 1.64 [1.58,1.67] & 1.63 [1.59,1.67] & 1.64 [1.58,1.67] \\
\hline
\end{tabular}
\caption{\footnotesize{Non-linear Gaussian model: medians and $1^{st}-3^{rd}$ quartiles for estimates obtained on 30 independent simulations. (*)The IF2 method resamples at every time point, while SAEM-ABC and SAEM-SMC resamples only when $ESS<\bar{M}$. }} 
\label{tab:nonlingauss}
\end{table}

We then perform 30 simulations with SAEM-SMC using the same simulated data and parameters starting values as for SAEM-ABC. As from Table \ref{tab:nonlingauss} simulations for $\sigma_y$ converge to completely wrong values. For this case we also experimented with $M=5,000$ using $\bar{M}=2,000$ but this does not solve the problem with SAEM-SMC, even if we let the algorithm start at the true parameter values. We noted that when using $M=2,000$ particles with SAEM-ABC and $\bar{M}=200$ the algorithm resamples every fourth observation, and in a generic iteration we observed an ESS of about 100 at the last time point. Under the same setup SAEM-SMC resampled at each time point and resulted in an ESS of about 10 at the last time point (a study on the implications of frequent resampling is considered in section \ref{sec:impoverishment}). Therefore we now perform a further simulation study to verify whether using a smaller $\bar{M}$ (hence reducing the number of times resampling is performed) can improve the performance of SAEM-SMC. Indeed, using for example $\bar{M}=20$ gives better results for SAEM-SMC, see Table \ref{tab:nonlingauss} (there we only use $M=1,000$ for comparison between methods). This is further investigated in section \ref{sec:impoverishment}.

\subsubsection{Comparison with iterative filtering and a pseudo-marginal Bayesian algorithm}

\begin{figure}
\centering
\includegraphics[scale=0.5]{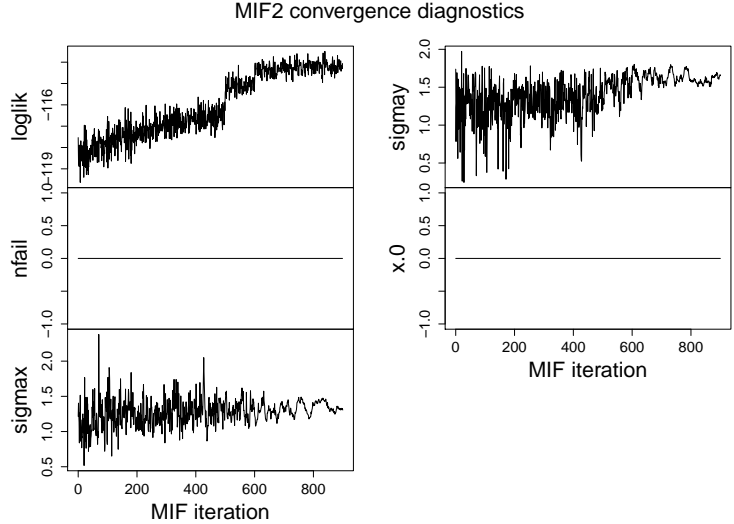}
\caption{\footnotesize{Non-linear Gaussian model: traces obtained with IF2 when using $M=1,000$ particles. (Top left) evolution of the loglikelihood function; (bottom left) evolution of $\sigma_x$; (top right) evolution of $\sigma_y$.}}
\label{fig:nonlingauss-IF2}
\end{figure}

We compare the results above with the iterated filtering methodology for maximum likelihood estimation (IF2, \citealp{ionides2015inference}), using the R package \texttt{pomp} \citep{king2015statistical}. We provide \texttt{pomp} with the same data and starting parameter values as considered in SAEM-ABC and SAEM-SMC. We do not provide a detailed description of IF2 here: it suffices to say that in IF2 particles are generated for both parameters $\btheta$ (e.g. via perturbations using random walks) and for the systems state (using the bootstrap filter). Moreover, same as with ABC methods, a ``temperature'' parameter (to use an analogy with the simulated annealing optimization method) is let decrease until the algorithm ``freezes'' around an approximated MLE. This temperature parameter, here denoted with $\epsilon$, is decreased in $\epsilon\in\{0.9,0.7,0.4,0.3,0.2\}$ which seems appropriate as explained below, where the first value is used for the first 500 iterations of IF2, then   each of the remaining values is used for 100 iterations, for a total of 900 iterations. Notice that the tested version of \texttt{pomp} (v. 1.4.1.1) uses a bootstrap filter that resamples at every time point, hence results obtained with IF2 are not directly comparable with SAEM-ABC and SAEM-SMC. Results are in Table \ref{tab:nonlingauss} and a sample output from one of the simulations obtained with $M=1,000$ is in Figure \ref{fig:nonlingauss-IF2}. From the loglikelihood in  Figure \ref{fig:nonlingauss-IF2} we notice that the last major improvement in likelihood maximization takes place  at iteration 600 when $\epsilon$ becomes $\epsilon=0.7$. Reducing $\epsilon$ further does not give any additional benefit (we have verified this in a number of experiments with this model). Notice that in order to run, say, 400 iterations of IF2 with $M=1,000$ for a \textit{single} experiment, instead of thirty, it required about 70 seconds. That is IF2 is about fourteen times slower than SAEM-ABC, although the comparison is not completely objective as we coded SAEM-ABC with Matlab, while IF2 is provided in an R package with forward model simulation implemented in C.

Finally we use a particle marginal method (PMM, \citealp{andrieu2009pseudo}) on a single simulation (instead of thirty), as this is a fully Bayesian methodology and results are not directly comparable with SAEM nor IF2. The PMM we construct approximates the likelihood function of the state space model using a bootstrap filter with $M$ particles, then plugs such likelihood approximation into a Metropolis-Hastings algorithm. As remarkably shown in \cite{beaumont2003estimation} and \cite{andrieu2009pseudo}, a PMM returns a Markov chain for $\btheta$ having the posterior $\pi(\btheta|\bY_{1:n})$ as its stationary distribution. This implies that PMM is an algorithm producing exact Bayesian inference for $\btheta$, regardless the number of particles used to approximate the likelihood.

Once more we make use of facilities provided in \texttt{pomp} to run PMM. We set uniform priors $U(0.1,15)$ for both $\sigma_x$ and $\sigma_y$ and run 4,000 MCMC iterations of PMM using 2,000 particles for the likelihood approximation. Also, we set the PMM algorithm in the most favourable way, by starting it at the true parameter values (we are only interested in the inference results, rather than showing the performance of the algorithm when starting from remote values). 
The proposal function for the parameters uses an adaptive MCMC algorithm based on Gaussian random walks, and was tuned to achieve the optimal 7\% acceptance rate \citep{sherlock2015efficiency}. For this single simulation, we obtained the following posterior means and 95\% intervals: $\hat{ \sigma}_x=1.52$ [0.42,2.56], $\hat{ \sigma}_y=1.53$ [0.36,2.34].

\subsubsection{The particles impoverishment problem}\label{sec:impoverishment}

\begin{figure}
\centering
\includegraphics[scale=0.35]{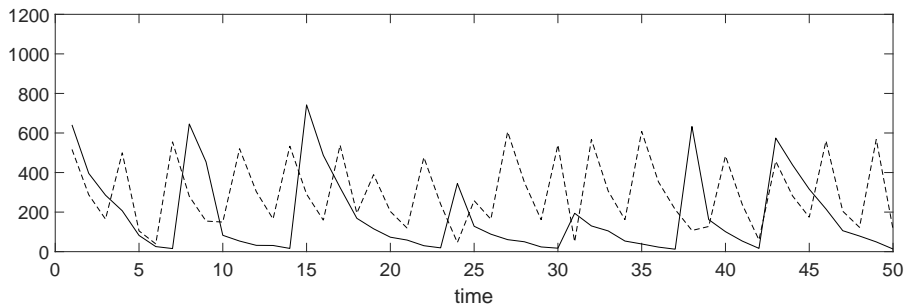}
\caption{Nonlinear-Gaussian model: ESS for SAEM-SMC when $(\bar{M},M)=(20,1000)$ (solid line) and $(\bar{M},M)=(200,1000)$ (dashed line) as a function of time.}\label{fig:ess-comparison}
\end{figure}

Figure \ref{fig:ess-comparison} reports the effective sample size ESS as a function of time $t$ (at a generic iteration of SAEM-SMC, the 20th in this case). As expected, for a smaller value of $\bar{M}$ the ESS is most of times smaller than when a larger $\bar{M}$ is chosen, with the exception of a few peaks. This is a direct consequence of performing resampling more frequently when $\bar{M}$ is larger. However, a phenomenon that is known to be strictly linked to resampling is that of ``samples impoverishment'', that is the resampling step reduces the ``variety'' of particles, by duplicating the ones with larger weight and killing the others. In fact, when many particles have a common ``parent'' at time $t$, these are likely to end close to each other at time $t+1$.
This has a negative impact on the inference because the purpose of the particles is to approximate the density \eqref{eq:filtering-density} (or \eqref{eq:abc-posterior}) which generates the trajectory sampled at step 2 in algorithms \ref{alg:saem-abc-smc} and \ref{alg:saem-smc}. Lack of variety in the particles reduces the quality of this approximation.

\begin{figure}
\centering
\includegraphics[scale=0.35]{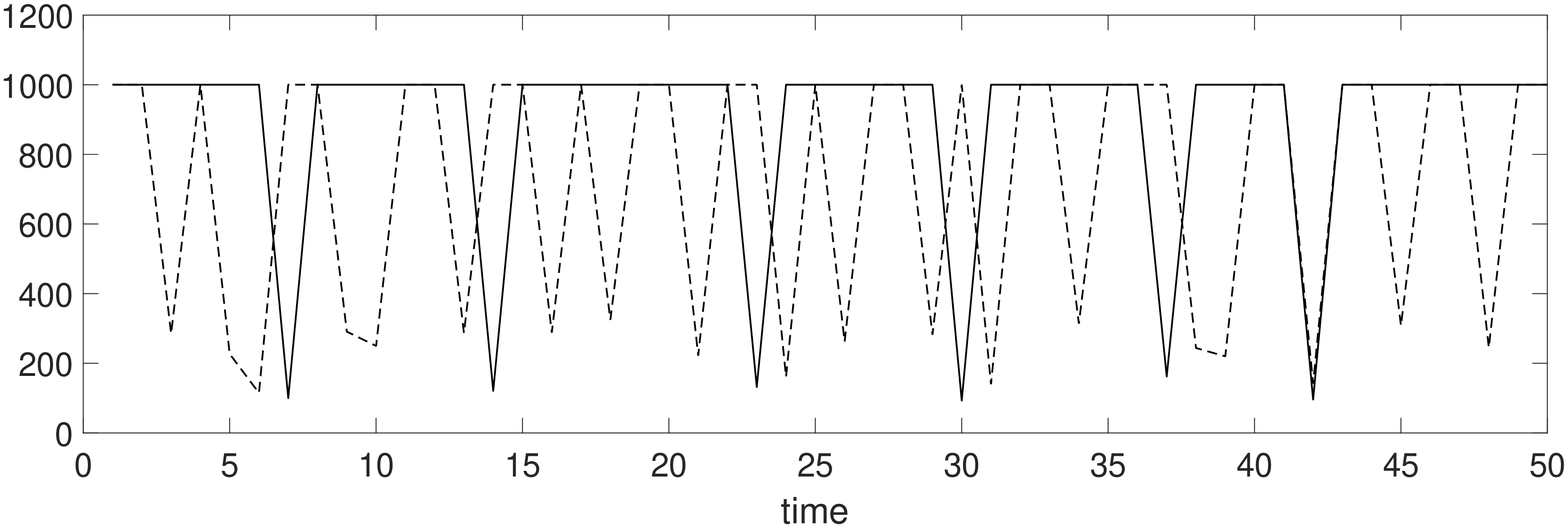}
\caption{Nonlinear-Gaussian model: number of distinct particles for SAEM-SMC when $(\bar{M},M)=(20,1000)$ (solid line) and $(\bar{M},M)=(200,1000)$ (dashed line) as a function of time.}\label{fig:distinct-particles-comparison}
\end{figure}

Indeed, Figure \ref{fig:distinct-particles-comparison} shows that the variety of the particles (as measured by the number of distinct particles) gets impoverished for a larger $\bar{M}$, notice for example that the solid line in Figure \ref{fig:distinct-particles-comparison} almost always reaches its maximum attainable value $M=1,000$, that is all particles are distinct, while this is often not the case for the dashed line. Since the trajectory $\bX^{(k)}$ that is selected at iteration $k$ of SAEM, either in step 2 of algorithm 3 or in step  2 of algorithm 4, follows from backward-tracing the genealogy of a certain particle, having variety in the cloud of particles is crucial here.  

\begin{figure}
\centering
\includegraphics[scale=0.35]{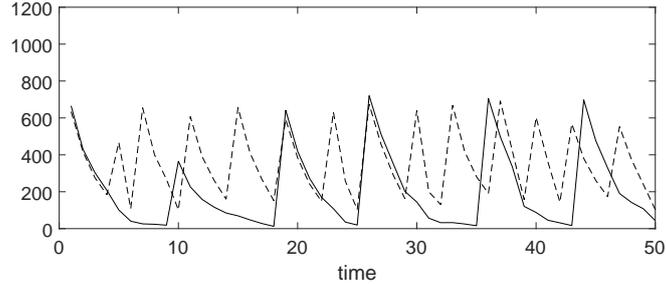}
\caption{Nonlinear-Gaussian model: ESS for SAEM-ABC when $(\bar{M},M)=(20,1000)$ (solid line) and $(\bar{M},M)=(200,1000)$ (dashed line) as a function of time.}\label{fig:ess-ABC-comparison}
\end{figure}

\begin{figure}
\centering
\includegraphics[scale=0.35]{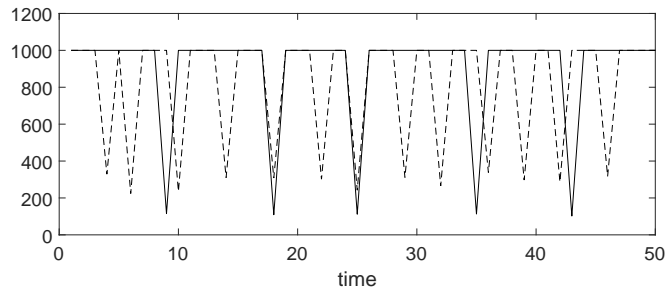}
\caption{Nonlinear-Gaussian model: number of distinct particles for SAEM-ABC when $(\bar{M},M)=(20,1000)$ (solid line) and $(\bar{M},M)=(200,1000)$ (dashed line) as a function of time.}\label{fig:distinct-particles-ABC-comparison}
\end{figure}

This seems related to the counter-intuitive good performance of the ABC-filter, even though SAEM-ABC is based on the additional approximation induced by the $J$ function in equation \eqref{eq:gauss-kernel}. We now produce plots for the ESS values and the number of distinct particles at the smallest value of the ABC threshold $\delta$. As we see in Figure \ref{fig:ess-ABC-comparison}, while the ESS are not much different from the ones in Figure \ref{fig:ess-comparison}, instead the number of distinct particles in Figure \ref{fig:distinct-particles-ABC-comparison} is definitely higher than the SAEM-SMC counterpart in Figure \ref{fig:distinct-particles-comparison}, meaning that such number drops below the maximum $M=1,000$ fewer times. For example from Figure \ref{fig:distinct-particles-comparison} we can see that when $(\bar{M},M)=(200,1000)$ the number of distinct particles drops 19 times  away from the maximum $M=1,000$, when using SAEM-SMC. With SAEM-ABC this number drops only 13 times (Figure \ref{fig:distinct-particles-ABC-comparison}). When the number of resampling steps is reduced, using $(\bar{M},M)=(20,1000)$, we have more even results, with the number of distinct particles dropping six times for SAEM-SMC and five times for SAEM-ABC.

As a support to this remark, refer to Table \ref{tab:meaness-meannumparticles}: we perform thirty independent estimation procedures, and in each we obtain the sample mean of the ESS (means computed over varying time $t$) and the sample mean of the number of distinct particles (again over varying $t$). Then, we report the mean over the thirty estimated sample means of the ESS (and corresponding standard deviation) and the same for the number of distinct particles. Clearly numbers are favourable to SAEM-ABC, showing consistently larger values (with small variation between the thirty experiments).

We argue that when many particles, as generated in the bootstrap filter, fail to get close to the target observations, then resampling is frequently triggered, this degrading the variety of the particles. However with an ABC filter, particles receive some additional weighting (due to a $J_{j,\delta}>1$ in \eqref{eq:gauss-kernel} when $Y_j^{*(m)}\approx Y_j$ and a small $\delta$) which allows for a larger number of particles to have a  non-negligible weight. While it is \textit{not true} that an ABC filter is in general expected to perform better than a non-ABC one, here we find that the naive bootstrap filter performs worse than the ABC counterpart for the reasons discussed above. 

\begin{table}
\small
\centering
\begin{tabular}{lll|ll}
\hline
\\
{} & \multicolumn{2}{c}{$(M,\bar{M})=(1000,20)$} & \multicolumn{2}{c}{$(M,\bar{M})=(1000,200)$} \\
\hline
{} &         {ESS}      &  \# distinct particles    & {ESS} & \# distinct particles\\
SAEM-SMC &  202.20 (61.70) & 871.83 (90.62) & 252.10 (83.31) & 616.40 (212.61)\\
SAEM-ABC &   202.62 (12.31) & 905.64 (7.95)  & 351.80 (14.75) & 812.85 (4.68)\\
\hline
\end{tabular}
\caption{\footnotesize{Nonlinear-Gaussian model: mean ESS and mean number of distinct particles (and corresponding standard deviations in parentheses) at a generic iteration of SAEM-ABC and SAEM-SMC. Averages and standard deviations are taken over 30 independent repetitions of the experiment.}}
\label{tab:meaness-meannumparticles}
\end{table}

\subsubsection{Relation with Gibbs sampling}\label{sec:gibbs}
As previously mentioned, the generation of $\bs_k$ in \eqref{eq:summaries-simulation} followed by corresponding parameter estimates $\hat{\btheta}^{(k)}$ in \eqref{eq:m-step} is akin to two steps of a Gibbs sampling algorithm, with the important distinction that in SAEM $\hat{\btheta}^{(k)}$ is produced by a deterministic step (for given $\bs_k$). We show that the construction of a Gibbs-within-Metropolis sampler is possible, but that a naive implementation fails while a ``non-central parametrization'' seems necessary \citep{papaspiliopoulos2007general}. For given initial vectors $(\bX^{(0)},\sigma^{{(0)}}_x,\sigma^{{(0)}}_y)$, we alternate sampling from the conditional distributions (i) $p(\bX^{(b)}|\sigma^{{(b-1)}}_x,\sigma^{{(b-1)}}_y,\bY)$, (ii) $p(\sigma^{{(b)}}_x|,\sigma^{{(b-1)}}_y,\bX^{(b)},\bY)$ and (iii)  $p(\sigma^{{(b)}}_y|,\sigma^{{(b)}}_x,\bX^{(b)},\bY)$ where $b$ represents the iteration counter in the Gibbs sampler ($b=1,2,...$). The resulting multivariate sample $(\bX^{(b)},\sigma^{{(b)}}_x,\sigma^{{(b)}}_y)$ is a draw from the posterior distribution $\pi(\bX,\sigma_x,\sigma_y|\bY)$. We cannot sample from the conditional densities in (i)--(iii), however at a generic iteration $b$ it is possible to incorporate a single Metropolis-Hastings step targeting the corresponding densities in (i)--(iii) separately, resulting in a Metropolis-within-Gibbs sampler,  see e.g. \cite{liu2008monte}. Notice that what (i) implies is a joint sampling (block-update) for all the elements in $\bX$, however it is also possible to sample individual elements one at a time (single-site update) from $p(X^{(b)}_j|X^{(b)}_{j-1},X^{(b-1)}_{j+1},\sigma^{{(b-1)}}_x,\sigma^{{(b-1)}}_y,\bY)$. For both single-site and block-update sampling, the mixing of the resulting chain is very poor, this resulting from the high correlations between sampled quantities, notably the correlation between elements in $\bX$ and also between $\bX$ and $(\sigma_x,\sigma_y)$. However, there exists a simple solution based on breaking down the dependence between some of the involved quantities: for example at iteration $b$ we propose in block a vector $\bX^\#$ generated ``blindly'' (i.e. conditionally on the current parameter values $(\sigma^{{(b-1)}}_x,\sigma^{{(b-1)}}_y)$, but unconditionally on $\bY$) by iterating through \eqref{eq:nonlingauss-model} then accept/reject the proposal according to a Metropolis-Hastings step, then define $\bX^{*(b)}:=\tilde{\bX}/\sigma_x^{(b-1)}$, where $\tilde{\bX}$ is the last accepted proposal of $\bX$ (that is $\tilde{\bX}\equiv \bX^\#$ if $\bX^\#$ has been accepted). Sample from (ii) $p(\sigma^{{(b)}}_x|,\sigma^{{(b-1)}}_y,\bX^{*(b)},\bY)$ and (iii)  $p(\sigma^{{(b)}}_y|,\sigma^{{(b)}}_x,\bX^{*(b)},\bY)$, and transform back to $\tilde{\bX}:=\sigma_x^{(b)}\bX^{*(b)}$. This type of updating scheme is known as non-central parametrization \citep{papaspiliopoulos2007general}. The expressions for the conditional densities in (i)--(iii) are given in the appendix for the interested reader. Note that the SAEM-ABC and SAEM-SMC do not require a non-central parametrization. The maximization step of SAEM smooths out the correlation between the proposed parameter and the latent state, and the numerical convergence of the algorithms still occur (this has also been noticed in previous papers on SAEM-MCMC,  see \citealp{kuhn2005maximum, donnet2008parametric}). 

\begin{figure}
\centering
\includegraphics[width=17cm,height=4cm]{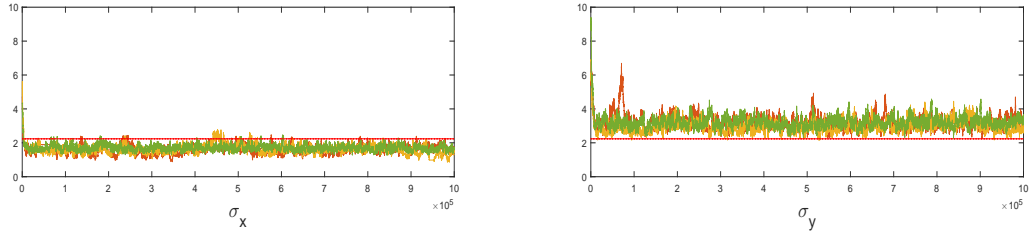}
\caption{\footnotesize{Nonlinear Gaussian model: three chains with different starting values from the Metropolis-within-Gibbs sampler for data generated with $(\sigma_x,\sigma_y)=(2.23,2.23)$. Horizontal lines are the true parameter values.}}
\label{fig:nonlingauss-gibbs-traces}
\end{figure}

\begin{figure}
\centering
\includegraphics[width=13cm,height=3.5cm]{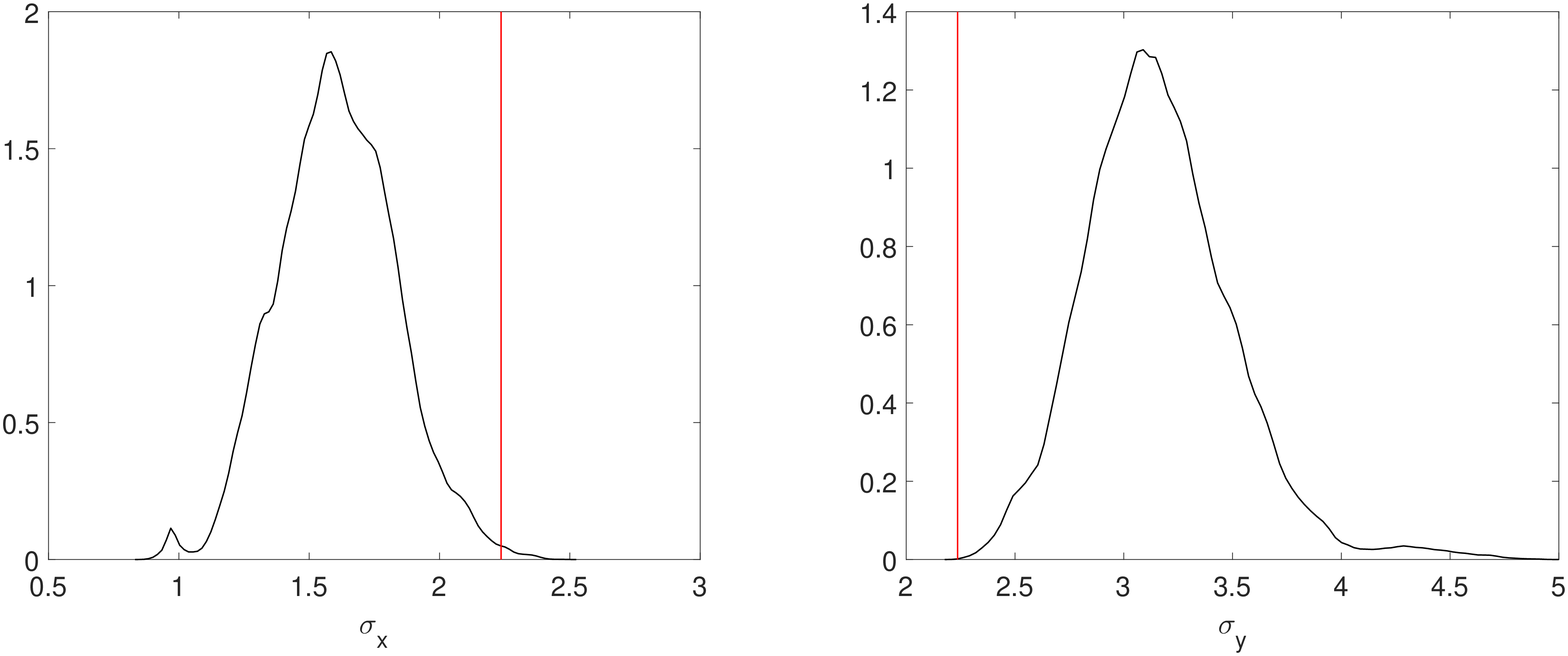}
\caption{\footnotesize{Nonlinear Gaussian model: marginal posteriors from the Metropolis-within-Gibbs sampler for data generated with $(\sigma_x,\sigma_y)=(2.23,2.23)$. Vertical lines are the true parameter values.}}
\label{fig:nonlingauss-gibbs-marginals}
\end{figure}

Once more, we attempt estimating data produced with $(\sigma_x,\sigma_y)=(2.23,2.23)$ (as in section \ref{sec:nonlingauss-largenoise}) this time using Metropolis-within-Gibbs. Trace plots in Figures \ref{fig:nonlingauss-gibbs-traces} show satisfactory mixing for three chains starting at three random values around $(\sigma_x,\sigma_y)=(6,6)$. However, while the chains initialized at values far from the truth rapidly approach the true values, the 95\% posterior intervals fail to include them, see Figure \ref{fig:nonlingauss-gibbs-marginals}. If we re-execute the same experiment, this time with data generated with smaller noise $(\sigma_x,\sigma_y)=(0.71,0.71)$, it seems impossible to catch the ground truth parameters. From a typical chain, we obtain the following posterior means and 95\% posterior intervals: $\hat{\sigma}_x=1.59$ [1.18,2.02],  $\hat{\sigma}_y=1.63$ [1.28,2.08]. Clearly for noisy time-dependent data, such as state-space models, particles based methods seem to better address cases where noisy data are affected by both measurement and systemic noise. 

\subsection{A pharmacokinetics model}\label{sec:theoph}

Here we consider a model for pharmacokinetics dynamics. For example we could formulate a model to study the Theophylline drug pharmacokinetics. This example has often been described in literature devoted to longitudinal data modelling with random parameters (mixed--effects models), see \cite{pinheiro1995approximations} and \cite{donnet2008parametric}. Same as in \cite{picchini2014inference} here we do not consider a mixed--effects model.
We denote with $X_t$ the level of drug concentration in blood at time $t$ (hrs).
Consider the following non-authonomous stochastic differential equation:
\begin{equation}
dX_t = \biggl(\frac{Dose\cdot K_a \cdot K_e}{Cl}e^{-K_a t}-K_eX_t\biggr)dt + \sigma \sqrt{X_t} dW_t, \qquad t\geq t_0
\label{eq:theophylline-sde}
\end{equation}
where $Dose$ is the known drug oral dose received by a subject, $K_e$ is the elimination rate constant, $K_a$ the absorption rate constant, $Cl$ the clearance of the drug and $\sigma$ the intensity of the intrinsic stochastic noise.
We simulate data at $n=100$ equispaced sampling times $\{t_1,t_{\Delta},...,t_{100\Delta}\}=\{1,2,...,100\}$ where $\Delta=t_j-t_{j-1}=1$. The drug oral dose is chosen to be 4 mg.
After the drug is administered, we consider as $t_0=0$ the time when the concentration first reaches $X_{t_0}=X_0=8$. The error model is assumed to be linear, $Y_j=X_j+\varepsilon_j$ where the $\varepsilon_j\sim N(0,\sigma_{\varepsilon}^2)$ are i.i.d., $j=1,...,n$. Inference is based on data $\{Y_1,...,Y_{n}\}$ collected at corresponding sampling times. Parameter $K_a$ is assumed known, hence parameters of interest are $\btheta=(K_e,Cl,\sigma^2,\sigma_{\varepsilon}^2)$ as $X_0$ is also assumed known.

Equation \eqref{eq:theophylline-sde} has no available closed-form solution, hence simulated data are created in the following way. We first simulate numerically a solution to \eqref{eq:theophylline-sde} using the Euler--Maruyama discretization with stepsize $h=0.05$ on the time interval $[t_0,100]$. The Euler-Maruyama scheme is given by
\[
X_{t+h} = X_{t} + \biggl(\frac{Dose\cdot K_a \cdot K_e}{Cl}e^{-K_a t}-K_eX_t\biggr)h + (\sigma \sqrt{h\cdot X_t})Z_{t+h}
\]
where the $\{Z_t\}$ are i.i.d. $N(0,1)$ distributed. The grid of generated values $\bX_{0:N}$ is then linearly interpolated at sampling times $\{t_1,...,t_{100}\}$ to give $\bX_{1:n}$. Finally residual error is added to $\bX_{1:n}$ according to the model $Y_j=X_j+\varepsilon_j$ as explained above. Since the errors $\varepsilon_j$ are independent, data $\{Y_j\}$ are conditionally independent given the latent process $\{X_t\}$.

\paragraph{Sufficient statistics for SAEM}

The complete likelihood is given by
\[
 p(\bY,\bX_{0:N};\btheta)=p(\bY|\bX_{0:N};\btheta)p(\bX_{0:N};\btheta)=\prod_{j=1}^n p(Y_j|X_j;\btheta)\prod_{i=1}^Np(X_i|X_{i-1};\btheta)
\]
where the unconditional density $p(x_0)$ is disregarded in the last product since we assume $X_0$ deterministic. Hence the complete-data loglikelihood is
\[
L_c(\bY,\bX_{0:N};\btheta) =\sum_{j=1}^n \log p(Y_j|X_j;\btheta)+\sum_{i=1}^N \log p(X_i|X_{i-1};\btheta).
\]
Here $p(y_j|x_j;\btheta)$ is a Gaussian with mean $x_j$ and variance $\sigma^2_{\varepsilon}$. The transition density $p(x_i|x_{i-1};\theta)$ is not known for this problem, hence we approximate it with the Gaussian density induced by the Euler-Maruyama scheme, that is
\begin{equation}
p(x_i|x_{i-1};\btheta)\approx \frac{1}{\sigma\sqrt{2\pi x_{i-1}h}}\exp\biggl\{-\frac{\bigl[x_i-x_{i-1}-(\frac{Dose\cdot K_a \cdot K_e}{Cl}e^{-K_a \tau_{i-1}}-K_ex_{i-1})h\bigr]^2}{2\sigma^2x_{i-1}h}\biggr\}. \label{eq:euler-gaussdensity}
\end{equation}
Notice the Gaussian distribution implied by \eqref{eq:euler-gaussdensity} shares some connection with tools developed for optimal states predictions in signal processing, such as the unscented Kalman filter, e.g. \cite{sitz2002estimation}.
We now derive sufficient summary statistics for the parameters of interest, based on the complete loglikelihood.
Regarding $\sigma^2_{\varepsilon}$ this is trivial as we only have to consider $\sum_{j=1}^n \log p(y_j|x_j;\theta)$ to find that a sufficient statistic is $S_{\sigma^2_{\varepsilon}}=\sum_{j=1}^n(y_j-x_j)^2$. Regarding the remaining parameters we have to consider $\sum_{i=1}^N\log p(x_i|x_{i-1};\btheta)$. For $\sigma^2$ a sufficient statistic is
\[
S_{\sigma^2} = \sum_{i=1}^N\biggl(\frac{\bigl[x_i-x_{i-1}-(\frac{Dose\cdot K_a \cdot K_e}{Cl}e^{-K_a \tau_{i-1}}-K_ex_{i-1})h\bigr]^2}{x_{i-1}h}\biggr).
\]
Regarding $K_e$ and $Cl$ reasoning is a bit more involved: we can write
\begin{align*}
\sum_{i=1}^N \log p(x_i|x_{i-1};\btheta)
&\propto  \sum_{i=1}^N \frac{\bigl[x_i-x_{i-1}-(\frac{Dose\cdot K_a \cdot K_e}{Cl}e^{-K_a \tau_{i-1}}-K_ex_{i-1})h\bigr]^2}{x_{i-1}}\\
&= \sum_{i=1}^N \biggl[\frac{x_i-x_{i-1}}{\sqrt{x_{i-1}}}-\biggl(\frac{Dose\cdot K_a \cdot K_e}{Cl{\sqrt{x_{i-1}}}}e^{-K_a \tau_{i-1}}-\frac{K_ex_{i-1}}{\sqrt{x_{i-1}}}\biggr)h\biggr]^2.
\end{align*}
The last equality suggests a linear regression approach $E(V)=\beta_1C_1+\beta_2C_2$ for ``responses'' $V_i=(x_i-x_{i-1})/\sqrt{x_{i-1}}$ and ``covariates'' 
\begin{align*}
C_{i1}&=\frac{Dose\cdot K_a e^{-K_a\tau_{i-1}}h}{\sqrt{x_{i-1}}}\\
C_{i2}&= -\frac{x_{i-1}}{\sqrt{x_{i-1}}}h=-\sqrt{x_{i-1}}h
\end{align*}
and $\beta_1=K_e/Cl$, $\beta_2=K_e$. By considering the design matrix $\bC$ with columns $\bC_1$ and $\bC_2$, that is $\bC = [\bC_1, \bC_2]$, from standard regression theory we have that $\hat{\boldsymbol{\beta}}=(\bC'\bC)^{-1}\bC'\bV$ is a sufficient statistic for $\boldsymbol{\beta}=(\beta_1,\beta_2)$, where $'$ denotes transposition. We take $S_{K_e}:=\hat{\beta}_2$ also to be used as the updated value of $K_e$ in the maximisations step of SAEM. Then we have that $\hat{\beta}_1$ is sufficient for the ratio $K_e/Cl$ and use $\hat{\beta}_2/\hat{\beta}_1$ as the update of $Cl$ in the M-step of SAEM. The updated values of $\sigma$ and $\sigma_{\varepsilon}$ are given by $\sqrt{S_{\sigma^2}/N}$ and $\sqrt{S_{\sigma^2_{\varepsilon}}/n}$ respectively.

\subsection{Results}\label{ex:theophylline-results}

\begin{figure}
\centering
\includegraphics[width=18cm,height=9cm]{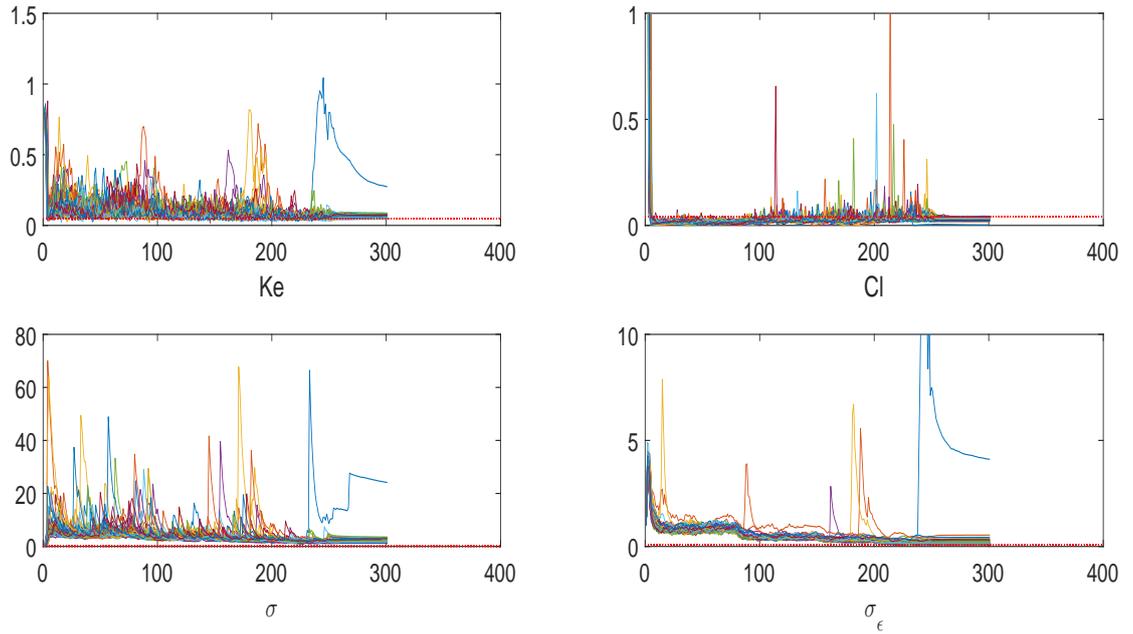} 
\caption{\footnotesize{Theophylline model: 50 independent estimations using $K=300$ iterations of SAEM-ABC when $M=200$ and $\delta\in\{0.5, 0.2, 0.1, 0.03\}$. Top: $K_e$ (left) and $Cl$ (right). Bottom: $\sigma$ (left) and $\sigma_{\epsilon}$ (right). Horizontal lines are the true parameter values.}}
\label{fig:theophylline-SAEM-ABC_smallerror_M200-iter}
\end{figure}

We consider an experiment where 50 datasets of length $n=100$ each are independently simulated using parameter values $(K_e,K_a,Cl,\sigma, \sigma_{\varepsilon})=(0.05,1.492,$ $0.04,0.1,0.1)$. All results pertaining SAEM-ABC use the Gaussian kernel \eqref{eq:gauss-kernel}. For SAEM-ABC we use a number of ``schedules'' to decrease the threshold $\delta$. One of our attempts decreases the threshold as $\delta\in\{0.5, 0.2, 0.1, 0.03\}$ (results for this schedule are reported as SAEM-ABC(1) in Table \ref{tab:theophylline}): same as in the previous application, all we need is to determine an appropriate order of magnitude for the largest $\delta$. As an heuristic, we have that the empirical standard deviation of the differences $|Y_{j}-Y_{j-1}|$ is about 0.3. The first value of $\delta$ is used for the first 80 iterations then it is progressively decreased every 50 iterations. For both SAEM-ABC and SAEM-SMC we use $K_1=250$ and $K=300$ and optimization started at parameter values very far from the true values: starting values are $K_e=0.80$, $Cl=10$, $\sigma=0.14$ and $\sigma_\varepsilon=1$.
We first show results using $(M,\bar{M})=(200,10)$. We start with SAEM-ABC, see Figure \ref{fig:theophylline-SAEM-ABC_smallerror_M200-iter} where the effect of using a decreasing $\delta$ is evident, especially on the trajectories for $\sigma_\varepsilon$. All trajectories but a single erratic one converge towards the true parameter values. Estimation results are in Table \ref{tab:theophylline}, giving results for three different decreasing schedules for $\delta$, reported as SAEM-ABC(0) for $\delta\in\{0.5, 0.2, 0.1, 0.05, 0.01\}$, the already mentioned SAEM-ABC(1) having $\delta\in\{0.5, 0.2, 0.1, 0.03\}$ and finally SAEM-ABC(2) where $\delta\in\{1, 0.4, 0.1\}$. Results are overall satisfactory for all parameters but $\sigma$, which remains unidentified for all attempted SAEM algorithms, including those discussed later.
The benefit of decreasing $\delta$ to values close to zero are noticeable, that is among the algorithms using ABC, SAEM-ABC(0) gives the best results. 

\begin{figure}
\centering
\includegraphics[width=18cm,height=9cm]{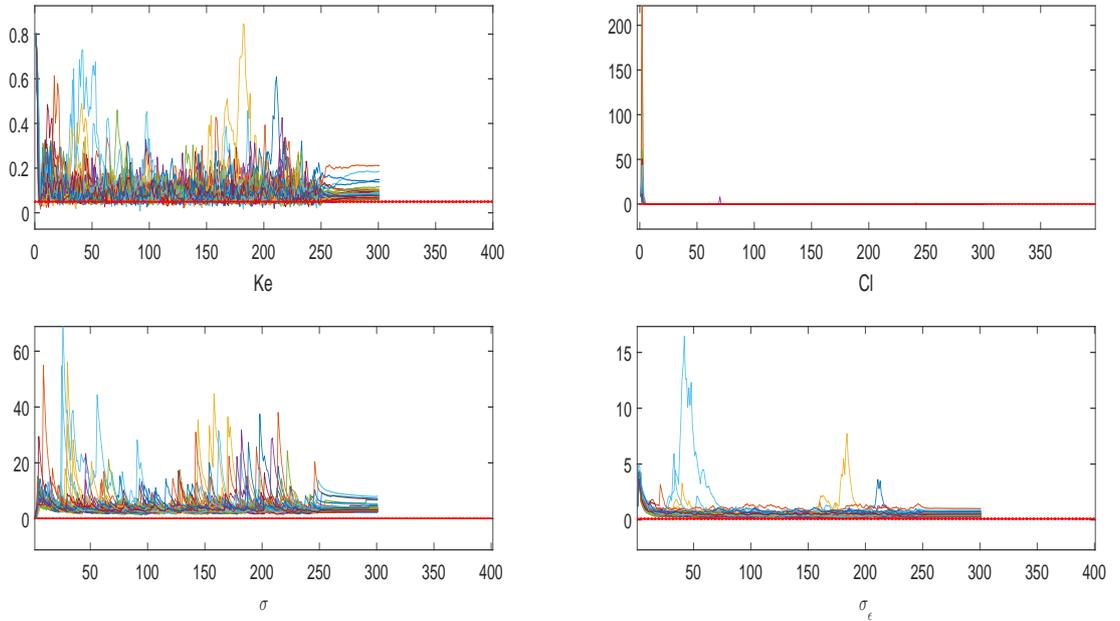} 
\caption{\footnotesize{Theophylline model: 50 independent estimations using $K=300$ iterations of SAEM-SMC when $M=200$. Top: $K_e$ (left) and $Cl$ (right). Bottom: $\sigma$ (left) and $\sigma_{\epsilon}$ (right). Horizontal lines are the true parameter values.}}
\label{fig:theophylline-SAEM-SMC_smallerror_M200-iter}
\end{figure}

\begin{figure}
     \subfloat[SAEM-ABC(0) at $t=10$.]{%
       \includegraphics[width=0.5\textwidth]{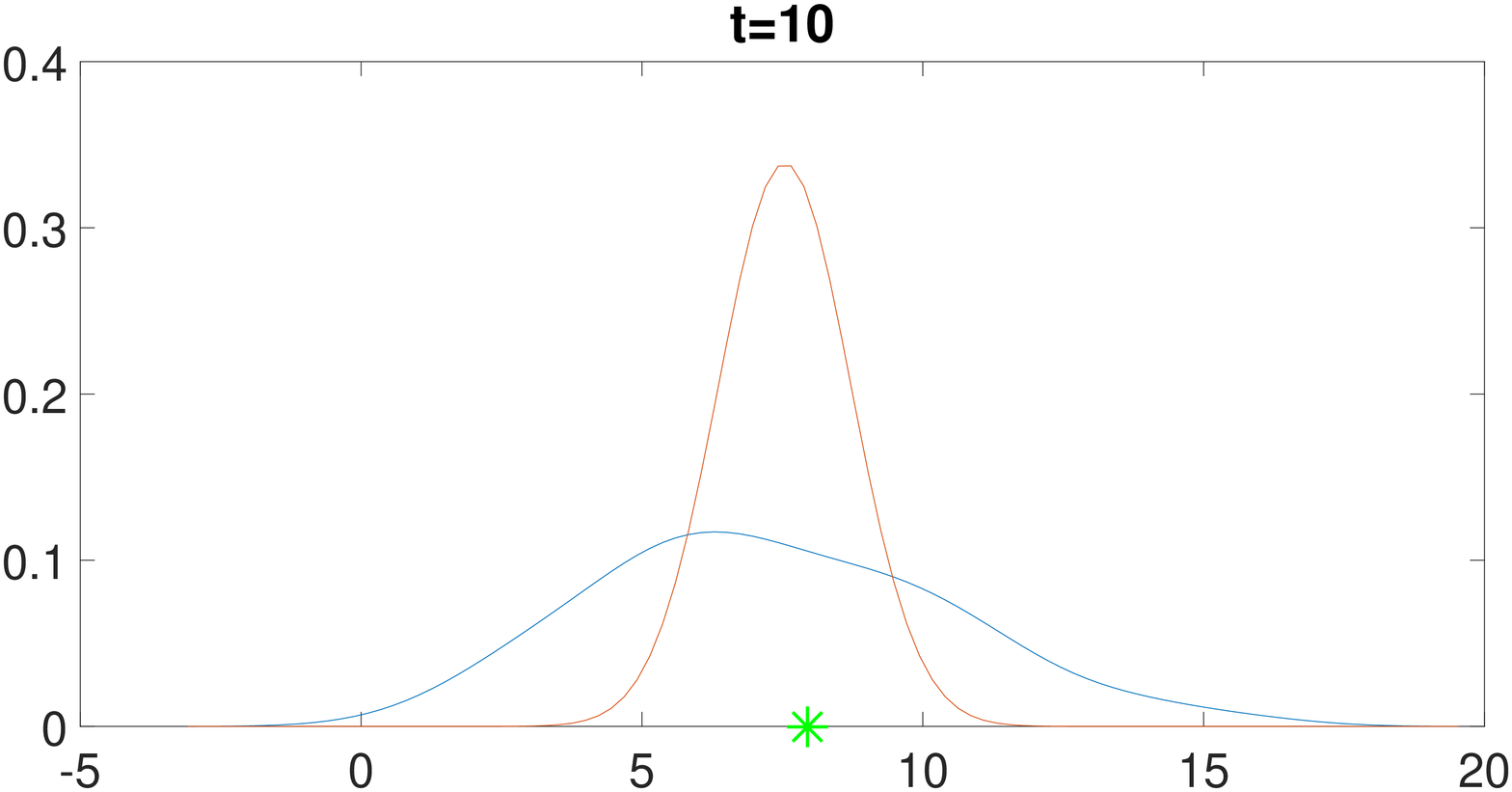}
     }
     \hfill
     \subfloat[SAEM-SMC at $t=10$.]{%
       \includegraphics[width=0.5\textwidth]{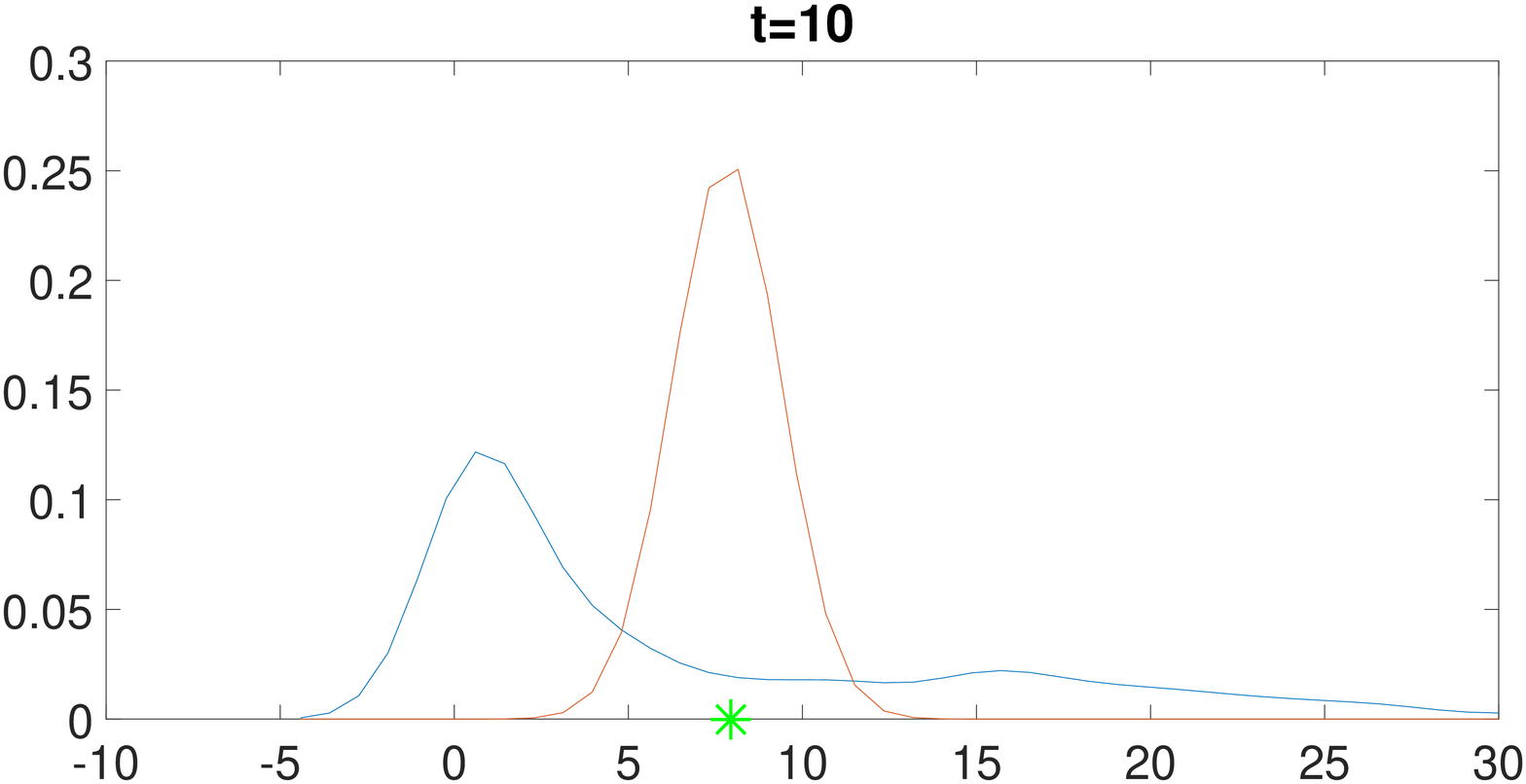}
     }\\
     \subfloat[SAEM-ABC(0) at $t=40$.]{%
       \includegraphics[width=0.5\textwidth]{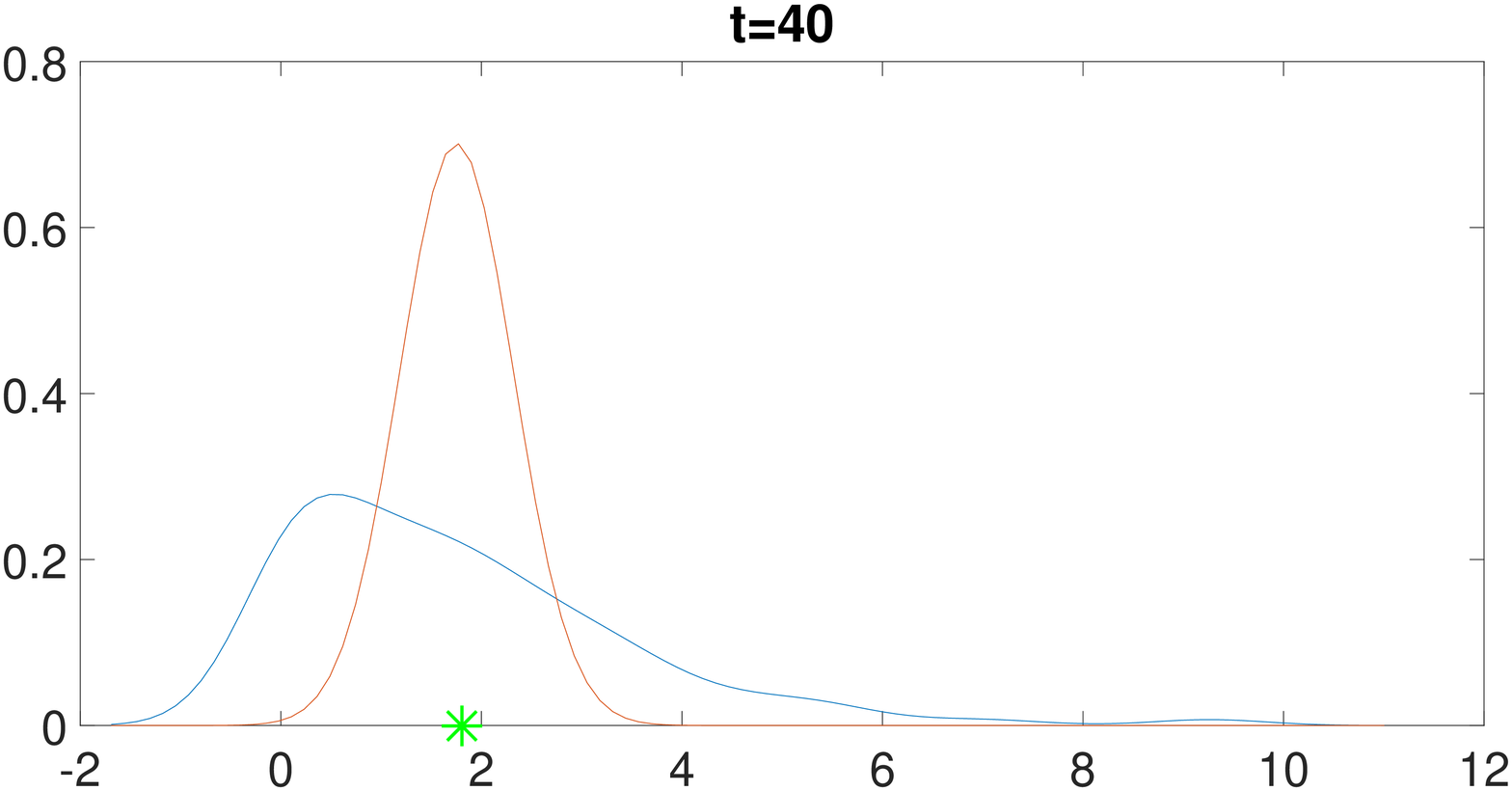}
     }
     \hfill
     \subfloat[SAEM-SMC at $t=40$.]{%
       \includegraphics[height=3.8cm,width=8cm]{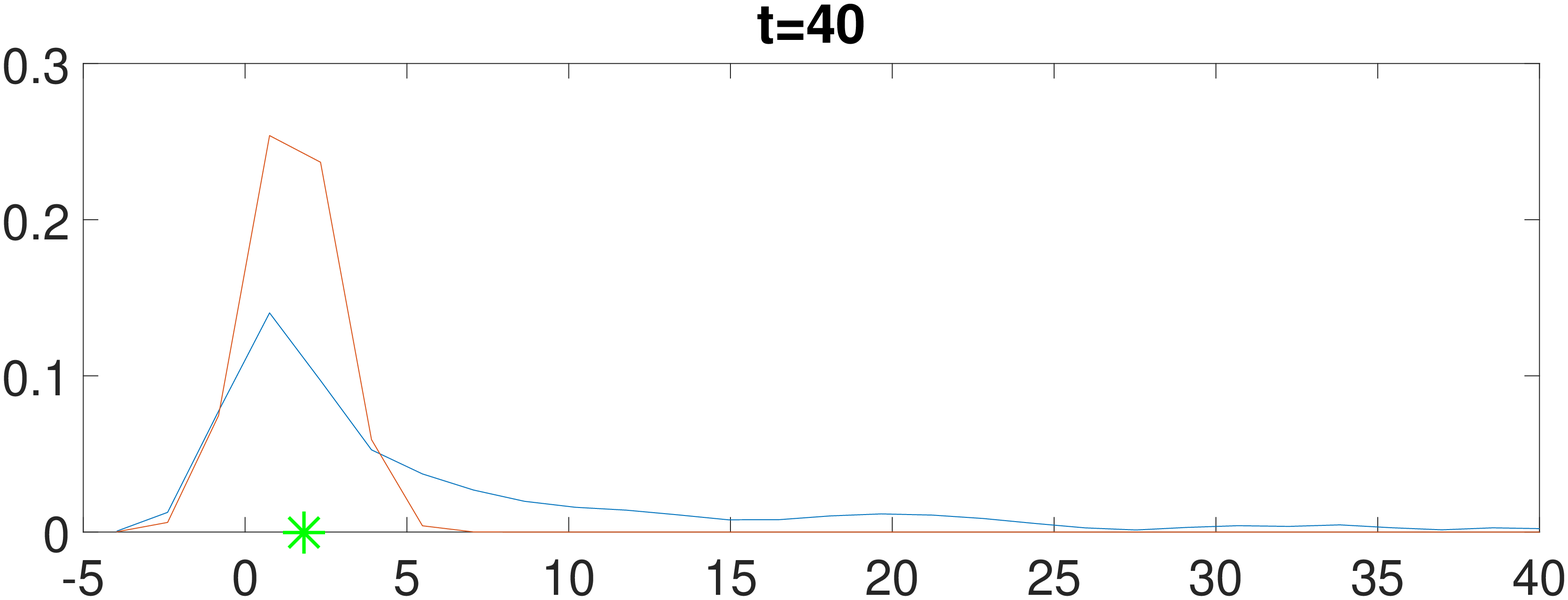}
     }\\
      \subfloat[SAEM-ABC(0) at $t=70$.]{%
       \includegraphics[width=0.5\textwidth]{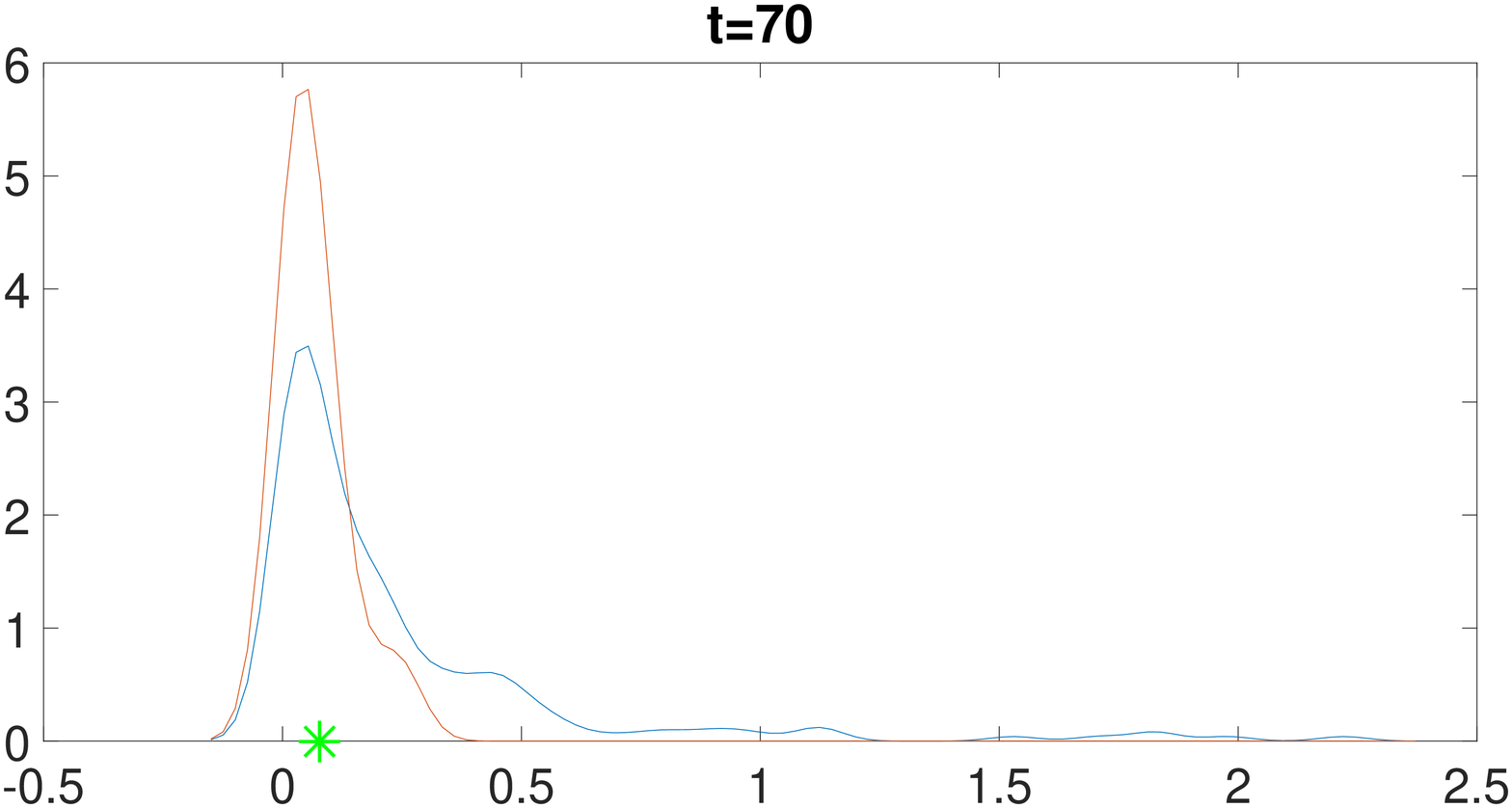}
     }
     \hfill
     \subfloat[SAEM-SMC at $t=70$.]{%
       \includegraphics[width=0.5\textwidth]{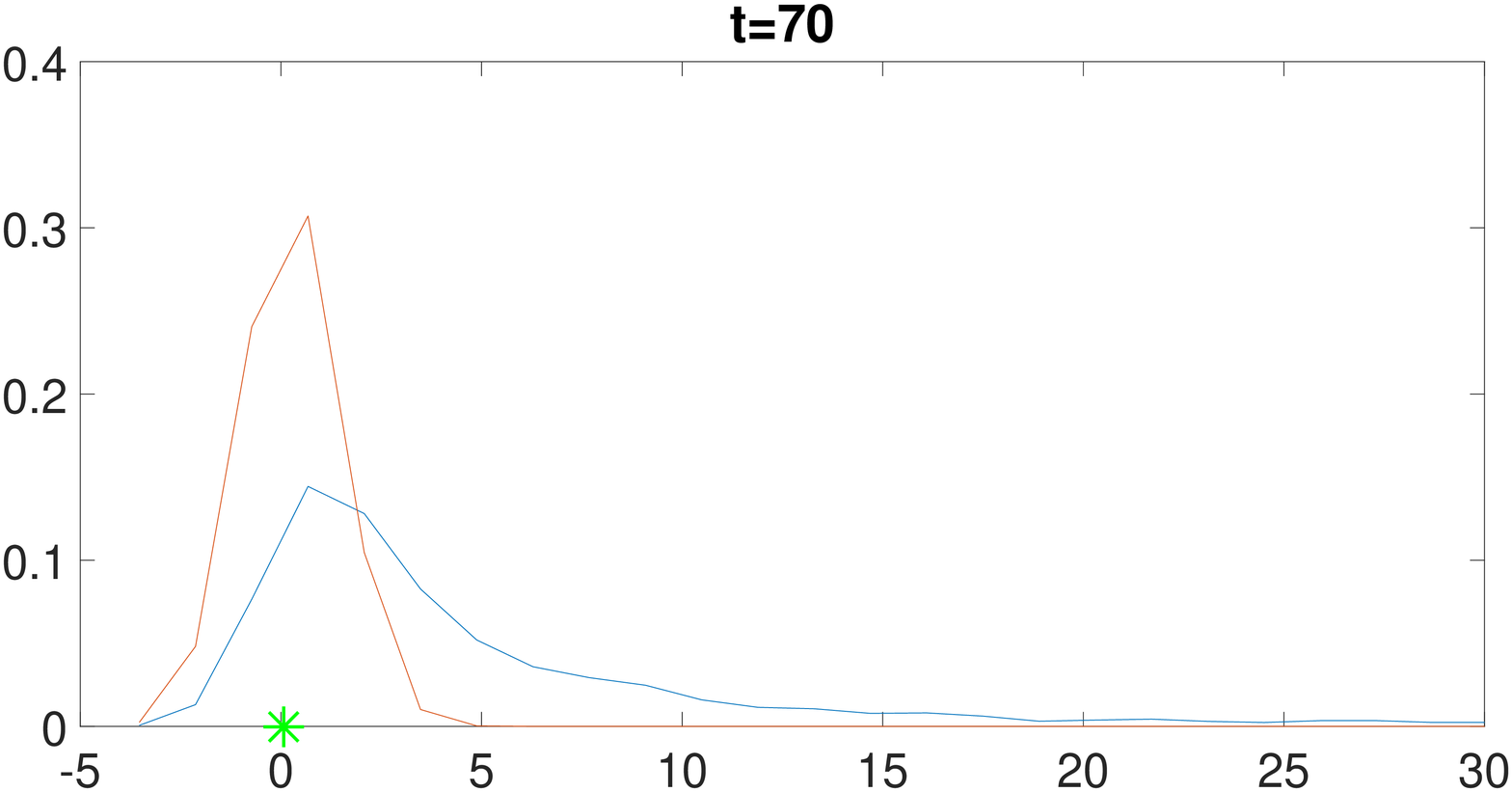}
     }
     \caption{\footnotesize{Kernel smoothed approximations of $\pi(\bX_{t}|\bY_{1:t-1})$ (blue) and $\pi(\bX_{t}|\bY_{1:t})$ (orange) at different values of $t$ for the case $(M=200,\bar{M}=10)$. Green asterisks are the true values of $\bX_t$. Notice scales on x-axes are different.}}
     \label{fig:theophylline-filteringdensities}
\end{figure}

From results in Table \ref{tab:theophylline} and Figure \ref{fig:theophylline-SAEM-SMC_smallerror_M200-iter}, again considering the case $M=200$, we notice that SAEM-SMC struggles in identifying most parameters with good precision. As discussed later on, when showing results with $M=30$, it is clear that when a very limited amount of particles is available (which is of interest when it is computationally demanding to forward-simulate from a complex model) SAEM-SMC is suboptimal compared to SAEM-ABC at least for the considered example. However results improve noticeably for SAEM-SMC as soon as the number of particles is enlarged, say to $M=1,000$. In fact for $M=1,000$ the inference results for SAEM-ABC(0) and SAEM-SMC are basically the same. It is of course interesting to uncover the reason why SAEM-ABC(0) performs better than SEM-SMC when the number of particles is small, e.g. $M=200$ (see later on for results with $M=30$).  In Figure \ref{fig:theophylline-filteringdensities} we compare approximations to the distribution $\pi(\bX_{t}|\bY_{1:t-1})$, that is the distribution of the state at time $t$ given previous data, and the filtering distribution $\pi(\bX_{t}|\bY_{1:t})$ including the most recent data. The former one, $\pi(\bX_{t}|\bY_{1:t-1})$, is approximated by kernel smoothing applied on the particles $\bX^{(m)}_t$. The filtering distribution $\pi(\bX_{t}|\bY_{1:t})$ is approximated by kernel smoothing assigning weights $W_t^{(m)}$ to the corresponding particles. It is from the (particles induced) discrete distribution approximating $\pi(\bX_{t}|\bY_{1:t})$ that particles are sampled when propagating to the next time point. Both densities are computed from particles obtained at the last SAEM iteration, $K$, for different values of $t$, at the beginning of the observational interval ($t=10$), at $t=40$ and towards the end of the observational interval ($t=70$). Since in Figure \ref{fig:theophylline-filteringdensities} we also report the true values of $\bX_t$, we can clearly see that, for SAEM-SMC, as $t$ increases the true value of $\bX_t$ is unlikely under $\pi(\bX_{t}|\bY_{1:t})$ (the orange curve). In particular, by looking at the orange curve in panel (f) in Figure \ref{fig:theophylline-filteringdensities} we notice that many of the particles ending-up far from the true $\bX_t=1.8$ receive non-negligible weight. Instead in panel (e) only particles very close to the true value of the state receive considerable weight (notice the different scales on the abscissas for panel (e) and (f)). Therefore it is not unlikely that for SAEM-SMC several ``remote'' particles are resampled and propagated. The ones resampled in SAEM-ABC received a weight $W_t^{(m)}\propto J_{t,\delta}$ which is large only if the simulated observation is very close to the actual observation, since $\delta$ is very small. 
Therefore for SAEM-SMC, when $M$ is small, the path sampled in step 2 of algorithm \ref{alg:saem-smc} (which is from $\pi(\bX_{1:t_n}|\bY_{1:t_n})$) is poor and the resulting inference biased. For a larger $M$ (e.g. $M=1,000$) SAEM-SMC enjoys a larger number of opportunities for particles to fall close to the targeted observation hence an improved inference.


\begin{table}
\small
\centering 
\begin{tabular}{lllll}
\hline\\
& \multicolumn{4}{c}{$(M,\bar{M})=(200,10)$} \\ 
 & $K_e$  &  $Cl$ & $\sigma$&  $\sigma_{\varepsilon}$\\ 
\hline
true values & 0.050& 0.040&   0.1& 0.1\\
\hline
SAEM-ABC(0) & 0.059 [0.054,0.067] & 0.034 [0.027,0.038] & 1.57 [1.35,2.15] & 0.15 [0.11,0.22]\\
SAEM-ABC(1) & 0.063 [0.058,0.074] & 0.031 [0.026,0.035] & 2.09 [1.56,2.80] & 0.16 [0.13,0.25]\\
SAEM-ABC(2) & 0.073 [0.065,0.088] & 0.025 [0.022,0.028] & 2.69 [2.44,3.50] & 0.32 [0.27,0.37]\\
SAEM-SMC&  0.078 [0.071,0.087] &   0.022 [0.019,0.027] &   3.45 [3.01,4.09] &   0.45 [0.36,0.55]\\
SAEM-GW & 0.061 [0.057,0.066]  &  0.032 [0.026,0.036] &   1.84 [1.44,2.81] & 0.12 [0.09,0.21] \\
\hline\\
& \multicolumn{4}{c}{$(M,\bar{M})=(1000,100)$} \\ 
 & $K_e$  &  $Cl$ & $\sigma$&  $\sigma_{\varepsilon}$\\ 
\hline
true values & 0.050& 0.040&   0.1& 0.1\\
\hline
SAEM-ABC(0) & 0.061 [0.054,0.065] & 0.033 [0.027,0.035] &  1.78 [1.48,2.48] & 0.13 [0.09,0.19]\\
SAEM-ABC(1) & 0.063 [0.059,0.070]   & 0.031 [0.025,0.035]  &  1.93 [1.65,2.41]   & 0.15 [0.13,0.23]\\
SAEM-ABC(2) & 0.073 [0.064,0.081] & 0.024 [0.021,0.028] & 2.98 [2.54,3.42] & 0.27 [0.24,0.35]\\
SAEM-SMC & 0.062 [0.057,0.068]  &  0.033 [0.027,0.036]  &  1.91 [1.40,2.25] &  0.12 [0.09,0.19]\\
SAEM-GW & 0.061 [0.057,0.068]  & 0.032 [0.025,0.035] &   1.90 [1.40,2.68] &    0.13 [0.1,0.24]\\
\hline
\end{tabular}     
\caption{\footnotesize{Theophylline: medians and $1^{st}-3^{rd}$ quartiles for estimates obtained on 50 independent simulations using SAEM-ABC   and SAEM-SMC. SAEM-ABC(0) denotes results obtained with $\delta\in\{0.5, 0.2, 0.1, 0.05, 0.01\}$, SAEM-ABC(1) denotes results obtained with $\delta\in\{0.5, 0.2, 0.1, 0.03\}$ and SAEM-ABC(2) denotes results obtained with $\delta\in\{1, 0.4, 0.1\}$.}}
\label{tab:theophylline}
\end{table}

An improvement over the bootstrap filter used in SAEM-SMC is given by a  methodology where particles are not proposed from the  transition density of the latent process (the latter proposes particles ``blindly'' with respect to the next data point). For example, \cite{golightly2011bayesian} consider a proposal distribution based on a diffusion bridge, conditionally to observed data. Their methodology is specific for state-space models driven by a stochastic differential equation whose approximate solution is obtained via the Euler-Maruyama discretisation, which is what we require. We use their approach to ``propagate forward'' particles. We write SAEM-GW to denote a SAEM algorithm using the proposal sampler in \cite{golightly2011bayesian}. By looking at Table \ref{tab:theophylline} we notice the improvement over the simpler SAEM-SMC when $M=200$. In fact SAEM-GW gives the best results of all SAEM-based algorithms we attempted, the downside being that such sampler is not very general and is only applicable to a specific class of models, namely (i) state-space models having an SDE that has to be numerically solved via Euler-Maruayama, (ii) observations having additive Gaussian noise, and (iii) observations having a state entering linearly, e.g. $y=a\cdot x+\varepsilon$ for some constant $a$. 

Same as in section \ref{sec:nonlingauss}, we consider Bayesian estimation using a particle marginal method (PMM). PMM is run with 1000 particles for 2,000 MCMC iterations. However it turns out that PMM cannot be initialized at the same remote starting values we used  for SAEM, as the approximated log-likelihood function at the starting parameter results not-finite for the considered number of particles. Therefore we let PMM start at $K_e=0.05$, $Cl=0.04$, $\sigma=0.2$, $\sigma_\varepsilon=0.3$ and use priors $K_e\sim U(0.01,0.2)$, $Cl\sim U(0.01,0.2)$, $\sigma\sim U(0.01,0.3)$, $\sigma_\varepsilon\sim U(0.05,0.5)$. Posterior means and 95\% intervals are: $K_e=0.076$ [0.067,0.084], $Cl=0.056$ [0.047,0.068], $\sigma=0.13$ [0.10,0.16], $\sigma_\varepsilon=0.12$ [0.099,0.139]. Hence, for starting parameter values close to the true values PMM behaves well (and estimates $\sigma$ correctly), but otherwise it might be impossible to initialize PMM, as also shown in \cite{fasiolo2016comparison}.

\begin{figure}
\centering
\includegraphics[width=11cm,height=5cm]{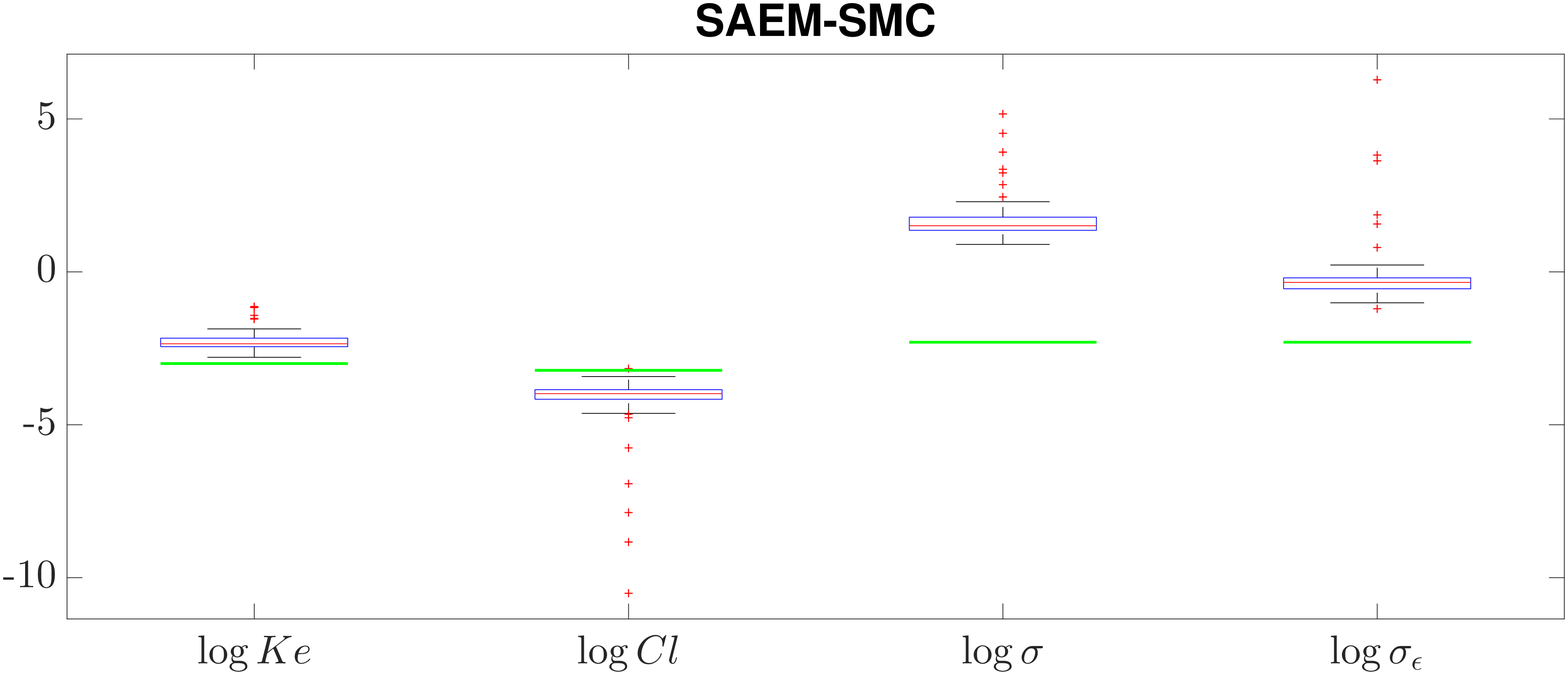} 
\caption{\footnotesize{Theophylline model: estimates obtained with SAEM-SMC when $M=30$. From left to right: $\log K_e$, $\log Cl$, $\log\sigma$ and $\log\sigma_\varepsilon$. Green lines are the true parameter values on log-scale.}}
\label{fig:theophylline_boxplots_SMC_smallerror_M30}
\end{figure}

\begin{figure}
\centering
\includegraphics[width=11cm,height=5cm]{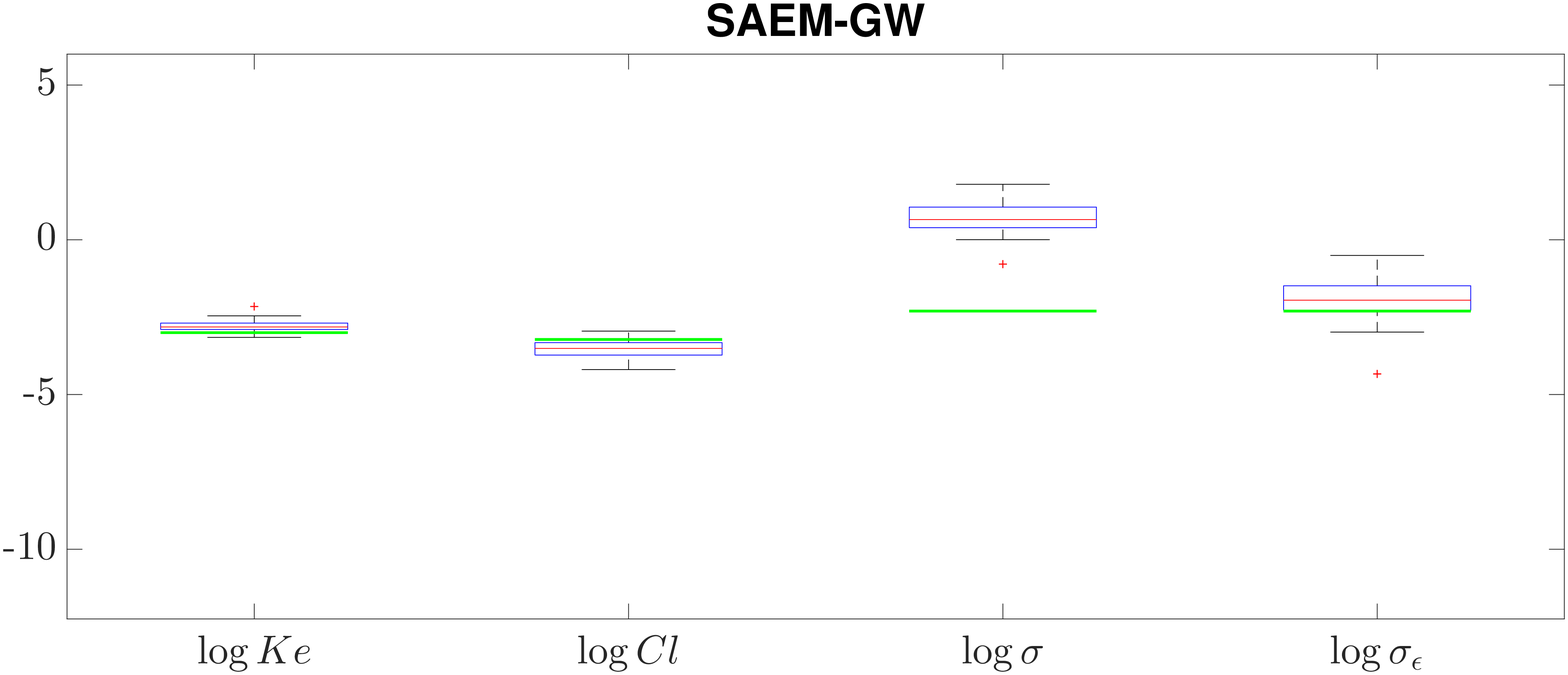} 
\caption{\footnotesize{Theophylline model: estimates obtained with the Golightly-Wilkinson sampler coupled with SAEM (SAEM-GW) when $M=30$. From left to right: $\log K_e$, $\log Cl$, $\log\sigma$ and $\log\sigma_\varepsilon$.  Green lines are the true parameter values on log-scale.}}
\label{fig:theophylline_boxplots_SMC-GW_smallerror_M30}
\end{figure}

\begin{figure}
\centering
\includegraphics[width=11cm,height=5cm]{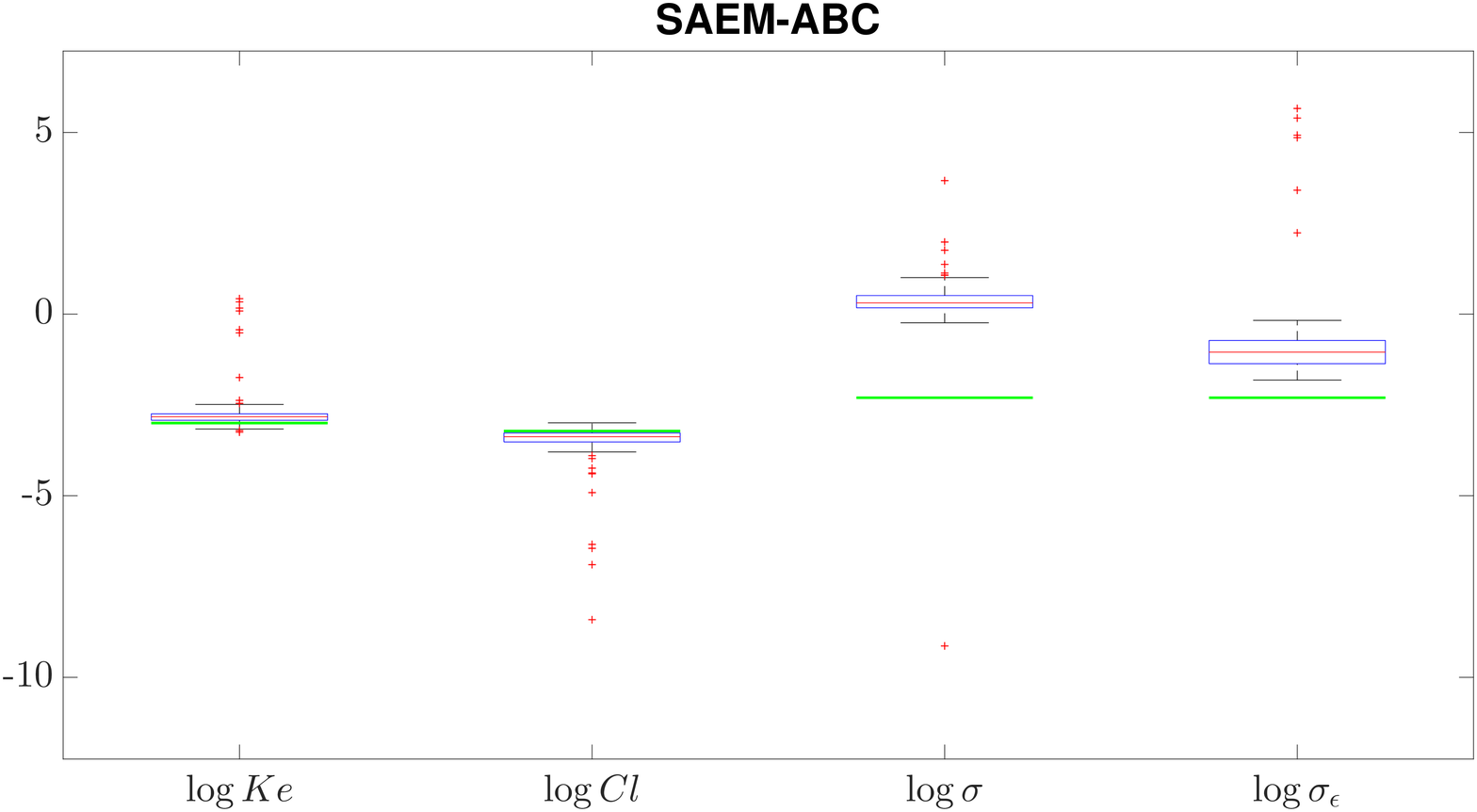} 
\caption{\footnotesize{Theophylline model: estimates obtained with SAEM-ABC when $M=30$. From left to right: $\log K_e$, $\log Cl$, $\log\sigma$ and $\log\sigma_\varepsilon$. Green lines are the true parameter values on log-scale.}}
\label{fig:theophylline_boxplots_ABC_smallerror_M30}
\end{figure}

We now explore whether the ABC approach can be of aid when saving computational time is essential, for example when simulating from the model is expensive. Although the model here considered can be simulated relatively quickly (it requires numerical integration hence computing times are affected by the size of the integration stepsize $h$), assuming this is not the case we explore what could happen if we can only afford running a simulation with $M=30$ particles. We run 100 simulations with this setting.  To ease graphic representation in the presence of outliers, estimates are reported on log-scales in the boxplots in Figures \ref{fig:theophylline_boxplots_SMC_smallerror_M30}--\ref{fig:theophylline_boxplots_ABC_smallerror_M30}. We notice that, with such a small number of particles, SAEM-ABC is still able to estimate $K_e$ and $Cl$ accurately whereas SAEM-SMC returns more biased estimates for all parameters. Also, while both methodologies fail in estimating $\sigma$ and $\sigma_\varepsilon$ when $M=30$ SAEM-ABC is still better than SAEM-SMC confirming the previous finding that, should the computational budget be very limited, SAEM-ABC is a viable option. SAEM-GW is instead able to estimate also the residual error variability $\sigma_\varepsilon$.

\section{Summary}\label{sec:summary}

We have introduced a methodology for approximate maximum likelihood estimation of the parameters in state-space models, incorporating an approximate Bayesian computation (ABC) strategy. The general framework is the stochastic approximation EM algorithm (SAEM) of \cite{Delyon1999}, and we embed a sequential Monte Carlo ABC filter into SAEM. SAEM requires  model-specific analytic computations, at the very least the derivation of sufficient statistics for the complete log-likelihood, to approach the parameters maximum likelihood estimate with minimal computational effort. However at any iteration of SAEM, it is required the availability of a filtered trajectory of the latent systems state, which we provide via the ABC filter of \cite{jasra2012filtering}. We call this algorithm SAEM-ABC. An advantage of using the ABC filter is its flexibility, as it is possible to modify its setup to
influence the weighting of the particles. In other words, it is possible to tune a positive tolerance $\delta$ to enhance the importance of those particles that are the closest to the observations. This produced sampled trajectories for step 2 of algorithm \ref{alg:saem-abc-smc} that resulted in a less biased parameter inference. This observation turned especially true for experiments run with a limited number of particles ($M=30$ or 200), which is relevant for computationaly intensive models not allowing for the propagation of a large number of particles. We compared our SAEM-ABC algorithm with a version of SAEM employing the bootstrap filter of \cite{gordon1993novel}. The bootstrap filter is the simplest sequential Monte Carlo algorithm, and is typically a default option in many software packages (e.g. the \texttt{smfsb} and \texttt{pomp} R packages, \citealp{smfsb} and \citealp{king2015statistical} respectively). In our work we show that, in some cases, SAEM-SMC (that is SAEM using a bootstrap filter) is sometimes inferior to SAEM-ABC, for example when the number of particles is small (section \ref{ex:theophylline-results}) or when too frequent resampling causes ``particles impoverishment'' (section \ref{sec:impoverishment}). 
SAEM-ABC requires the user to specify a sequence of ABC thresholds $\delta_1>\cdots \delta_L >0$. For one-dimensional time-series setting these thresholds is intuitive, since these represent standard deviations of perturbed (simulated) observations. Therefore the size of the largest one ($\delta_1$) can be determined by looking at plots of the observed time-series. 

Ultimately, while we are not claiming that an ABC filter should in general be preferred to a non-ABC filter, as the former one induces some approximation, it can be employed when it is difficult to construct (or implement) a more advanced sequential Monte Carlo filter. In our second application we consider a sequential Monte Carlo filter due to \cite{golightly2011bayesian}. This filter (which we call GW) is specifically designed for state-space models driven by stochastic differential equations (SDEs) requiring numerical discretization. While GW improves noticeably over the basic bootstrap filter, GW is not very general: again, it is specific for state-space models driven by SDEs;  the observation equation must have latent states entering linearly and  measurement errors must be Gaussian distributed. The ABC approach instead does not impose any limitation on the model structure.

\section*{Acknowledgements}
We thank anonymous reviewers for helping improve considerably the quality of the work. 
Umberto Picchini was partially funded by the Swedish Research Council (VR grant 2013-5167). Adeline Samson was partially supported by the LabEx PERSYVAL-Lab (ANR-11-LABX-0025-01) funded by the French program Investissement d’avenir.
This work is published on \textit{Computational Statistics}, \texttt{doi:10.1007/s00180-017-0770-y}.

\bibliographystyle{plainnat}  
\bibliography{biblio}

\appendix

\section*{Conditional densities for the Gibbs sampler in section \ref{sec:gibbs}.}
Here we report the conditional densities for the Gibbs sampler when $\bX$ is sampled in block. Here $\pi(\sigma_x)$ and $\pi(\sigma_y)$ are prior densities.

\begin{align*}
p(\sigma_x|\sigma_y,\bX,\bY)& \propto \prod_{j=1}^n \frac{1}{{\sigma_x}}e^{-\frac{1}{2\sigma^2_x}(X_j-2\sin(\exp(X_{j-1})))^2}\pi(\sigma_x)\\
p(\sigma_y|\sigma_x,\bX,\bY)& \propto \prod_{j=1}^n \frac{1}{{\sigma_y}}e^{-\frac{1}{2\sigma^2_y}(Y_j-X_j)^2}\pi(\sigma_y)\\
p(\bX|\sigma_x,\sigma_y,\bY)& \propto \prod_{j=1}^n \frac{1}{\sigma_x\sigma_y}e^{-\frac{1}{2\sigma^2_y}(Y_j-X_j)^2-\frac{1}{2\sigma^2_x}(X_j-2\sin(\exp(X_{j-1})))^2}
\end{align*}

\section*{First and second derivatives for the example in section \ref{sec:nonlingauss}.}

Here we report the first and second derivatives of the complete log-likelihood $L_c(\bX,\bY;\btheta)$ with respect to $\theta=(\sigma^2_x,\sigma^2_y)$.

\begin{align*}
\frac{\partial L_c(\bX,\bY;\btheta)}{\partial \sigma^2_x} &= -\frac{n}{2\sigma_x^2} + \frac{1}{2\sigma_x^4}\sum_{j=1}^n [X_j-2\sin(\exp(X_{j-1}))]^2, \\
\frac{\partial L_c(\bX,\bY;\btheta)}{\partial \sigma^2_y} &= -\frac{n}{2\sigma_y^2} + \frac{1}{2\sigma_y^4}\sum_{j=1}^n (Y_j-X_j)^2,\\
\frac{\partial^2 L_c(\bX,\bY;\btheta)}{\partial \sigma_x^2\partial \sigma_y^2} &= \frac{\partial^2 L_c(\bX,\bY;\btheta)}{\partial \sigma_y^2\partial \sigma_x^2} = 0,\\
\frac{\partial^2 L_c(\bX,\bY;\btheta)}{\partial (\sigma^2_x)^2} &= \frac{n}{2\sigma_x^4} -\frac{1}{\sigma_x^6} \sum_{j=1}^n [X_j-2\sin(\exp(X_{j-1}))]^2,\\ 
\frac{\partial^2 L_c(\bX,\bY;\btheta)}{\partial (\sigma^2_y)^2} &= \frac{n}{2\sigma_y^4} -\frac{1}{\sigma_y^6} \sum_{j=1}^n(Y_j-X_j)^2. 
\end{align*}

\section*{Fisher Information matrix  for the example in section \ref{sec:theoph}.}
To compute the Fisher Information matrix as suggested in section \ref{sec:fisher} we need to differentiate the complete data log-likelihood with respect to the four parameters $\btheta=(K_e, Cl, \sigma^2, \sigma_\varepsilon^2)$. We differentiate w.r.t. $(\sigma^2, \sigma_\varepsilon^2)$ instead of $(\sigma, \sigma_\varepsilon)$ because the complete log-likelihood is expressed as a function of sufficient statistics for $(\sigma^2, \sigma_\varepsilon^2)$.

Set, for $i=1, \ldots, N$,  $$z_i(\theta) = x_i-x_{i-1}-h( Dose\cdot K_a \cdot \frac{K_e}{Cl}\cdot e^{-K_a \tau_{i-1}}-K_e \cdot x_{i-1}).$$ 
 The four coordinates of the gradient are:
\begin{align*}
\frac{\partial}{\partial{K_e}} L_c(\bY, \bX; \btheta) &= -\frac1{\sigma^2}\sum_{i=1}^N \frac{z_i(\btheta)}{x_{i-1}}\biggl(x_{i-1}-\frac{Dose\cdot K_a}{Cl}e^{-K_a\tau_{i-1}}\biggr)\\
\frac{\partial}{\partial{Cl}} L_c(\bY, \bX; \btheta) &= -\frac1{\sigma^2}\sum_{i=1}^N \frac{z_i(\btheta)}{x_{i-1}} \biggl(\frac{Dose\cdot K_a\cdot K_e}{Cl^2}e^{-K_a\tau_{i-1}} \biggr)\\
\frac{\partial}{\partial{\sigma^2}} L_c(\bY, \bX; \btheta) &= - \frac N{2\sigma^2} +  \frac1{2h\sigma^4}\sum_{i=1}^N  \frac{z_i(\theta)^2}{x_{i-1}}\\
\frac{\partial}{\partial{\sigma_{\varepsilon}^2}} L_c(\bY, \bX; \theta) &= - \frac n{2\sigma_\varepsilon^2} +  \frac1{2\sigma_\varepsilon^4}\sum_{j=1}^n   (y_j-x_j)^2.\\
\end{align*}
Entries for the Fisher information matrix are (recall this is a symmetric matrix, therefore redundant terms are not reported. Further missing entries consist of zeros):
 \begin{eqnarray*}
\frac{\partial^2}{\partial^2{K_e}} L_c(\bY, \bX; \btheta) &=&- \frac h{\sigma^2}\sum_{i=1}^N (x_{i-1}-Dose\cdot \frac{K_a}{Cl} \cdot e^{-K_a \tau_{i-1}})^2\frac{1}{x_{i-1}}\\
\frac{\partial^2}{\partial^2{Cl}} L_c(\bY, \bX; \btheta) &=&
-\frac{1}{\sigma^2}\sum_{i=1}^N \biggl\{\frac{1}{x_{i-1}}\biggl[\frac{Dose\cdot K_a\cdot K_e}{Cl^2}e^{-K_a\tau_{i-1}}\biggl(h-\frac{2z_i(\btheta)}{Cl}\biggr)\biggr]\biggr\}\\
\frac{\partial^2}{\partial^2 {\sigma^2}} L_c(\bY, \bX; \btheta) &=&
\frac{N}{2\sigma^4}-\frac{1}{h\sigma^6}\sum_{i=1}^N \frac{z_i(\btheta)^2}{x_{i-1}}\\
\frac{\partial^2}{\partial^2 {\sigma^2_{\varepsilon}}} L_c(\bY, \bX; \btheta) &=& \frac{n}{2\sigma^4_{\varepsilon}}-\frac{1}{\sigma^{6}_\varepsilon}\sum_{j=1}^n (y_j-x_j)^2\\
\end{eqnarray*}
\begin{eqnarray*}
\frac{\partial^2}{\partial K_e\partial{Cl}} L_c(\bY, \bX; \btheta) &=& 
-\frac{1}{\sigma^2}\sum_{i=1}^N \frac{1}{x_{i-1}}\biggl\{\frac{Dose\cdot K_a}{Cl^2}e^{-K_a\tau_{i-1}}\biggl[h\cdot K_e\biggl(x_{i-1}-\frac{Dose\cdot K_a}{Cl}e^{-K_a\tau_{i-1}}\biggr)\\
&+&z_i(\btheta)\biggr]\biggr\}\\
\frac{\partial^2}{\partial \sigma^2\partial{K_e}} L_c(\bY, \bX; \btheta) &=&  \frac1{\sigma^4}\sum_{i=1}^N \frac{z_i(\btheta)}{x_{i-1}}\biggl(x_{i-1}-\frac{Dose\cdot K_a}{Cl}e^{-K_a\tau_{i-1}}\biggr)\\
\frac{\partial^2}{\partial \sigma^2\partial{Cl}} L_c(\bY, \bX; \btheta) &=&  \frac1{\sigma^4}\sum_{i=1}^N \frac{z_i(\btheta)}{x_{i-1}} \biggl(\frac{Dose\cdot K_a\cdot K_e}{Cl^2}e^{-K_a\tau_{i-1}} \biggr).
\end{eqnarray*}

\end{document}